\documentclass{aa}  
\usepackage{graphicx}
\usepackage{txfonts}
\usepackage{colortbl}

\begin{document}

\title{Astrophysical properties of 15062 Gaia DR3 gravity-mode pulsators}

\subtitle{Pulsation amplitudes, rotation, and spectral line
  broadening\thanks{The data file with the Gaia identification,
  fundamental parameters, dominant frequency and its amplitude, and
  the line broadening values for all 15062 stars in the studied sample
  are available electronically
at the CDS via anonymous ftp to cdsarc.cds.unistra.fr (130.79.128.5)
or via https://cdsarc.cds.unistra.fr/cgi-bin/qcat?J/A+A/???/A??}}

  \author{C.\ Aerts\inst{1,2,3,4}
\and G. Molenberghs\inst{5,6}
\and J. De Ridder\inst{1}
}

\institute{
Institute of Astronomy, KU Leuven, Celestijnenlaan 200D, B-3001 Leuven, Belgium\\
email:\ {Conny.Aerts@kuleuven.be}
\and
Department of Astrophysics, IMAPP, Radboud University Nijmegen, PO Box 9010,
6500 GL, Nijmegen, The Netherlands
\and 
Max Planck Institute for Astronomy, Koenigstuhl 17, 69117, Heidelberg, Germany
\and
Guest Researcher, Center for Computational Astrophysics, Flatiron Institute, 
    162 Fifth Ave, New York, NY 10010, USA
\and
I-BioStat, Universiteit Hasselt, Martelarenlaan 42, B-3500 Hasselt, Belgium
\and
I-BioStat, KU Leuven, Kapucijnenvoer 35, B-3000 Leuven, Belgium}

\date{Received December 17, 2022; Accepted February 8, 2023}

 
  \abstract
      {Gravito-inertial asteroseismology saw its birth thanks to
        high-precision CoRoT and {\it Kepler\/} space photometric
     light curves. So far, it gave rise to the internal rotation
     frequency of a few hundred intermediate-mass stars, yet only
     several tens of these have been weighed, sized, and
     age-dated with high precision from asteroseismic modelling.}
   {We aim to increase the sample of optimal targets for future
     gravito-inertial asteroseismology by assessing the properties of
     15062 newly found Gaia DR3 gravity-mode pulsators. 
     We also wish to investigate if there is any connection between their
     fundamental parameters and dominant mode on the one hand, and their
     spectral line broadening measured by Gaia on the other hand. }
   {After re-classifying about {  22}\% of the F-type gravity-mode
     pulsators as B-type according to their effective temperature, we construct
     histograms of the fundamental parameters and mode properties of
     the 15062 new Gaia DR3 pulsators. We compare these histograms with
     those of 63 {\it Kepler\/} bona fide class members. We fit 
     errors-in-variables regression models to 
     couple the effective temperature, luminosity, gravity, and
     oscillation properties to the two Gaia DR3 parameters capturing
     spectral line broadening for a fraction of the pulsators.  }
{We find that the selected 15062 gravity-mode pulsators have
  properties fully in line with those of their well-known {\it
    Kepler\/} analogues, revealing that Gaia has a role to play in
  asteroseismology.  The dominant g-mode frequency is a significant
  predictor of the spectral line broadening for the class members
  having this quantity measured. We show that the Gaia {\tt vbroad}
  parameter captures the joint effect of
  time-independent intrinsic and rotational line broadening and
  time-dependent tangential pulsational broadening. }
{While the Gaia mission was not designed to detect non-radial
  oscillation modes, its multitude of data and
  homogeneous data treatment allow us to identify a vast amount of new
  gravity-mode pulsators having fundamental parameters and dominant
  mode properties in agreement with those of such {\it Kepler\/} bona fide  
  pulsators. This large new sample of Gaia DR3 pulsators can be
  followed up with dedicated high-precision photometric or high-resolution
  spectroscopic instruments to embark upon asteroseismic modelling.}


\titlerunning{Astrophysical properties of Gaia DR3 gravity-mode pulsators}
\authorrunning{Aerts et al.}

\maketitle

%

\section{Introduction}

The Gaia space mission of the European Space Agency \citep{Prusti2016}
is currently revolutionising the entire field of astrophysics. Even
though Gaia is in the first place an astrometric mission, it also
delivers the largest homogeneous survey of broad-band photometric and
medium-resolution spectroscopic data \citep{Brown2016}.  While the
Gaia mission was not at all designed to deliver input for the research
field of asteroseismology \citep{Aerts2010}, it does contribute
important information for that recent emerging topic within stellar
astrophysics. Indeed, aside from stellar luminosities deduced from the
high-precision parallaxes \citep{BailerJones2018}, the Gaia
instrumentation also delivers years-long photometric time-series data
at milli-magnitude (mmag) precision in the Gaia G band. Although these
Gaia G light curves are only sparsely sampled, they do allow to
populate a wide range of stellar variability classes \citep[see][for a
  description of the ``variability tree'']{EyerMowlavi2008}. In
particular, Gaia data allows us to study the classes of pulsating
variables \citep[cf.,][Chapter\,2]{Aerts2010} with unprecendented
numbers of membership. \citet{Rimoldini2022} classified more than 12
million variables, among which RR\,Lyr stars \citep{Clementini2022},
Cepheids \citep{Ripepi2022}, young stellar objects
\citep{Marton2022}, and long period variables \citep{Lebzelter2022}.

In this work, we focus on stars observed by Gaia and classified from
its Data Release 3 \citep[DR3][]{Brown2021,Vallenari2022} as
gravity-mode (g-mode hereafter) pulsators by Coordination Unit\,7
treating variable stars \citep{Holl2018,Eyer2019,Eyer2022}. In their
Gaia DR3 Performance Verification Paper (PVP), \citet[][hereafter termed
  Paper\,I]{DeRidder2022} assigned the stars we revisit in the present
study to the classes of the Slowly Pulsating B \citep[SPB
  stars,][]{Waelkens1991,DeCatAerts2002} or $\gamma\,$Doradus stars
\citep[$\gamma\,$Dor,][]{Kaye1999,Handler1999}. These main-sequence
g-mode pulsators are the best laboratories for asteroseismic probing
of the deep interior of dwarfs with a mass between 1.3\,M$_\odot$ and 9\,M$_\odot$
\citep[][for a general review of asteroseismology of such
  stars]{Aerts2021}.  By now, hundreds of single and binary dwarfs with a convective
core have their rotation measured just outside their core
from series of consecutive radial-order dipole g-mode oscillations
\citep{Kurtz2014,Triana2015,Keen2015,Saio2015,VanReeth2016,Moravveji2016a,Murphy2016,Papics2017,Zwintz2017,VanReeth2018,GangLi2019,GangLi2020,Sekaran2021}.
Both the $\gamma\,$Dor and SPB pulsators reveal time-dependent spectral
line variations due to the tangential velocity fields at the
stellar surface
\citep{Aerts1999,DeCatAerts2002,Aerts2004,DeCat2006}. These pulsators
are intermediate-mass dwarfs in the core-hydrogen burning phase
without strong stellar winds.

The Gaia DR3 light curves analysed in Paper\,I resulted in an
order-of-magnitude increase in the population of the two classes of
non-radial g-mode main-sequence pulsators. The position of these new
candidate SPB and $\gamma\,$Dor pulsators in the Hertzsprung-Russell
diagram was compared with theoretically predicted instability strips,
{  each of which based on the dominant excitation mechanism for one particular
  choice of input physics, leading to}
coherent g~modes with infinite lifetime in Paper\,I. It was found
that many of the Gaia g-mode pulsators occur outside the borders of
{  such} instability strips for these two classes of g-mode pulsators. This
was ascribed to inaccuracies in the Gaia effective temperature, their
fast rotation and/or {  different input physics or}
(past) binarity not treated in instability
predictions, in addition to too low opacities of heavy elements such
as iron and nickel in the excitation layers as is well-known from
previous excitation studies of SPB pulsators
\citep{Moravveji2016b,Daszynska2017,Szewczuk2017}. Moreover, aside
from coherent eigenmodes with long lifetime driven by opacity layers
or at the bottom of the outer convective envelope, 
internal gravity waves with short lifetimes excited at the interface 
between the convective core and/or the convective outer
envelope and radiative zones have been suggested from
multi-dimensional hydrodynamical simulations mimicking dwarfs in the
considered mass regime
\citep{Rogers2013,Grassitelli2015a,Grassitelli2015b,Edelmann2019,Horst2020}.
All these predicted g~modes and internal gravity waves act together
in the stellar interior and those reaching the stellar surface with detectable amplitude give
rise to complex light curves and time-dependent spectral line profile
variations.
{  This is in agreement with modern time-resolved space photometric
  large surveys delivering $\mu$mag precision and highlighting a continuous coverage of observed  
intermediate-mass pulsating dwarfs along the main sequence 
\citep[][and
  Paper\,I]{Uytterhoeven2011,Bowman2019,Pedersen2019,Antoci2019,Bowman2020,Balona2020},
}
after earlier similar findings from high-resolution ground-based time-resolved
spectroscopy mentioned above.

Here, we study the astrophysical properties of the
new Gaia DR3 g-mode pulsators found in Paper\,I. We consider all these
pulsators having their luminosity, effective temperature, and
gravity determined by the Gaia Data Processing Analysis Consortium
\citep[DPAC,][]{Brown2016,Brown2021}. We compare the properties of
these Gaia DR3 g-mode pulsators with those of the sample of 63
best-known bona fide g-mode pulsators observed with the
NASA {\it Kepler\/} space telescope, for which high-resolution
follow-up spectroscopy was assembled and interpreted.  These {\it
  Kepler\/} and spectroscopic data led to the identification of dipole
g~modes of consecutive radial order
and to asteroseismic modelling for these 63 dwarfs
\citep{Mombarg2021,Pedersen2021}.

In the current follow-up study of Paper\,I we consider the amplitude
of the dominant frequency in the Gaia G band and relate it to the
fundamental parameters for both the Gaia DR3 and {\it Kepler\/} g-mode
pulsators. Moreover, we consider the sub-samples of Gaia DR3 g-mode
pulsators for which an estimation of the spectral line broadening is
available from Gaia's Radial Velocity Spectrometer (RVS) within the
large homogeneous Gaia DR3 data set \citep{Creevey2022,Fremat2022}.
Our general aim is to search for relationships between the fundamental
parameters and pulsational properties of g-mode pulsators. In
particular, we investigate if there is any connection between the
properties of the {  dominant g~mode of the} stars and their rotation and/or spectral
line broadening. So far, similar studies have been hampered by small
sample sizes \citep{Aerts2014a} or by separate and/or inhomogeneous
treatment of statistical modelling based on the observables deduced
from photometric and spectroscopic data
\citep{SSD2017,Burssens2020}. Even though the Gaia photometric and
spectroscopic instruments were not designed to study non-radial
oscillations nor optimised to deduce the line broadening of stars
hotter than 7\,000\,K, DR3 does provide unprecedentedly large samples
of homogeneously treated g-mode pulsators in terms of their line
broadening and fundamental parameters compared to those available in
the literature prior to Gaia DR3.

We discuss the sample selection for the current paper in Sect.\,2 and
consider the fundamental parameters and dominant variability
characteristics of 15062 g-mode pulsators in Sect.\,3 and Sect.\,4,
respectively.
Section\,5
focuses on the astrophysical interpretation of the measured spectral
line broadening of the sample, based on the method of
errors-in-variables and on multi-variable regression models constructed
via the technique of backward selection. We discuss our findings and
conclude in Sect.\,6.


\section{Sample selection}

Paper\,I resulted in 106\,207 Gaia DR3 main-sequence pulsators of
spectral types O, B, A, or F fullfilling four criteria: 1) their Gaia
DR3 G light curve consists of at least 40 epochs; 2) their
dominant cyclic frequency (denoted here as $\nu$) in the Gaia G light
curve occurs in the range $[0.7,25]$\,d$^{-1}$; 3) this frequency $\nu$
differs from any of the instrumental frequencies 4, 8, 12, 16, 20, and
24\,d$^{-1}$ by more than 0.05\,d$^{-1}$; 4) $\nu$ has a false alarm
probability in the definition by \citet{Baluev2008} below
0.001. Despite these already strict selection rules,
additional restrictions on the frequency interval to which $\nu$ had
to belong for each of the four considered pulsation classes were
imposed in Paper\,I to beat instrumental effects in Fourier space,
because they occur at mmag level and intervene with the signal of
non-radial oscillations also occuring at such level for dwarf stars of
intermediate mass.

The following four classes of pulsators were considered in Paper\,I:
$\beta\,$Cep stars, Slowly Pulsating B (SPB) stars, $\delta\,$Sct
stars, and $\gamma\,$Dor stars \citep[cf.\,][Chapter\,2]{Aerts2010}.
We refer to Paper\,I and its literature references for the details of
the additional selection rules imposed upon $\nu$ based on common
knowledge of the pulsational properties for these four well-known classes
of variables, but recall here that the dominant modes of $\beta\,$Cep
and $\delta\,$Sct stars are p~modes with observed frequencies
typically above 3\,d$^{-1}$, while SPB and $\gamma\,$Dor stars have
dominant g~modes with observed frequencies mostly below
3\,d$^{-1}$, except for the fastest rotators.

Within the sample of 106\,207 candidate pulsators assigned to the four
pulsation classes in Paper\,I, those having frequencies above the spin
frequency of the Gaia satellite are most affected by mmag-level
instrumental effects, which may lead to spurious frequencies unrelated
to the star. For this reason, we focus the current work on the two
classes of main-sequence pulsators having their dominant frequency
well below the 4\,d$^{-1}$ spin frequency of the Gaia satellite.  For
now, with the relatively sparse DR3 light curves, this is the best
approach to study the astrophysical properties of the Gaia DR3 g-mode
pulsators without being contaminated by spurious instrumental
frequencies.

Appendix\,B of Paper\,I discussed the results for the dominant
frequency $\nu$ in the Gaia DR3 light curves of the 63 bona fide
g-mode pulsators (26 SPB and 37 $\gamma\,$Dor stars) whose entire
amplitude spectrum is known with a level of precision better than
about $10^{-6}$\,d$^{-1}$ in cyclic frequency and a few $\mu$mag in amplitude
\citep{VanReeth2015,Pedersen2021}.  \citet{Aertsetal2021} relied on
the mode identification for all these stars' detected and identified
dipole modes of consecutive radial order to
deduce the convective and wave Rossby numbers for these best-known
{\it Kepler\/} g-mode pulsators, covering the mass range from
1.3\,M$_\odot$ up to about 9\,M$_\odot$. All these 63 pulsators have a
dominant Gaia G amplitude, $A_\nu$, below 35\,mmag and their dominant
frequency occurs in the interval $\nu\in [0.7,3.2]$\,d$^{-1}$ (see
Figs\,B.1 and B.2 in Paper\,I).  Some of the new g-mode pulsators
identified from Gaia DR3 in Paper\,I have higher dominant
frequencies. Moreover, some of the Gaia DR3 g-mode pulsators have
frequencies that are hard to unravel from the Gaia instrument
frequencies caused by the spinning of the satellite.

Guided by Figs.\,B.1 and B.2 in Paper\,I summarising the dominant
frequency and amplitude for the 63 bona fide {\it Kepler\/} g-mode
pulsators, we further apply a fifth and sixth constraint in this work, in
addition to the selecton criteria of Paper\,I mentioned above, namely
we demand that 5) $\nu\in [0.7,3.2]$\,d$^{-1}$ and 6) $A_\nu\leq
35\,$mmag. These two extra restrictions are applied to the SPB and
$\gamma\,$Dor stars assigned to those two g-mode pulsator classes in
Paper\,I. This is to ensure that we are dealing with non-radial
oscillations rather than satellite frequencies.  Moreover, we restrict
these two samples to those pulsators having a measurement of
$\log\,L$, $\log\,T_{\rm eff}$, and $\log\,g$ in the DR3 {\tt gspphot}
tables. We use those values in order to maximise the sample
size of g-mode pulsators treated in one homogeneous way by DPAC
routines, given that we need to cover temperatures
from $\sim\!6\,500\,$K all the way up to 25\,000\,K 
\citep[see Paper\,I for details
  and][]{Vallenari2022,Creevey2022}.

{  A continuous coverage of pulsating B, A, and F stars along the
  main sequence was found in Paper\,I, in agreement with {\it
    Kepler\/} and TESS results \citep[e.g.,][]{Balona2020}. Since the
  variability classification used in Paper\,I relied only on the Gaia
  G-band DR3 light curves, it cannot distinguish between B- and F-type
  pulsators without spectroscopic information
  \citep[cf.,][]{Pedersen2019,Gebruers2021}.  On the other
  hand, the {\it Kepler\/} data clearly revealed that $\gamma\,$Dor and SPB
  pulsators have different astrophysical and pulsational properties
  \citep{VanReeth2015,Saio2018,Pedersen2021}.  Thus, we wish to treat them as
  two separate classes. We do so by relying on the
  Gaia DR3 effective temperature to
  reclassify the g-mode pulsators. Following the upper
  limit in effective temperature from the instability predictions by
  \citet{Xiong2016} for $\gamma\,$Dor stars as treshold, we use $T_{\rm
    eff}=8500\,$K to distinguish between $\gamma\,$Dor and SPB
  candidates. In practice, we reclassify all $\gamma\,$Dor candidates
  as SPB if their effective temperature is above 8500\,K and, vice
  versa, we re-assign all SPB stars with a temperature below 8500\,K
  as $\gamma\,$Dor pulsator.  This leads to a re-classification of
  3244 $\gamma\,$Dor candidates as new SPB pulsators based on their
  Gaia DR3 effective temperature.  This re-assignment gives a
  fractional memberships of 29\% SPB and 71\% $\gamma\,$Dor pulsators,
  which is fully in line with a Salpeter-type initial mass function
  \citep[IMF,][]{Salpeter1955} for the typical masses of
  1.6\,M$_\odot$ for $\gamma\,$Dor stars \citep{Mombarg2021} and of
  4\,M$_\odot$ for SPB stars \citep{Pedersen2022a}. As we discuss later
  in the paper, another choice of the treshold temperature to
  distinguish the two classes does not impact any of the results. }

A critical aspect of the current study compared to other surveys
of g-mode pulsating dwarfs is that all the DR3 data and
observables were obtained in one homogeneous way in terms of data
selection and analysis. This is in contrast to the treatment of
ground-based photometry and spectroscopy obtained for the much smaller
dedicated asteroseismology samples for these two classes so far. While
Gaia DR3 can only deliver the dominant mode at this stage, it
provides by far the largest homogeneous survey of $\gamma\,$Dor and SPB
pulsators to date.

\section{Fundamental parameters of the g-mode pulsators in the two samples}

\begin{figure*}[th!]
\begin{center} 
\rotatebox{270}{\resizebox{6.1cm}{!}{\includegraphics{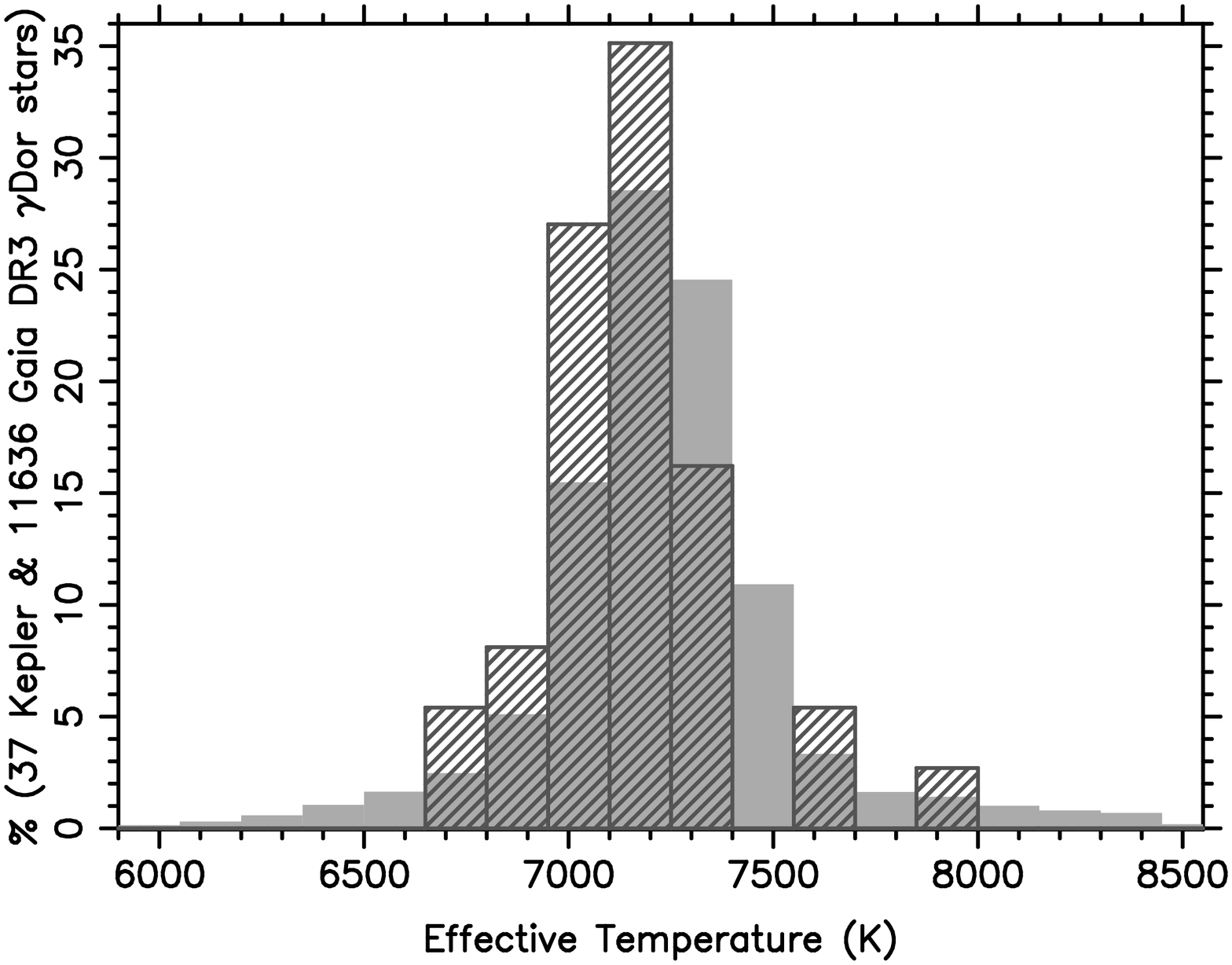}}}\hspace{0.5cm}
\rotatebox{270}{\resizebox{6.1cm}{!}{\includegraphics{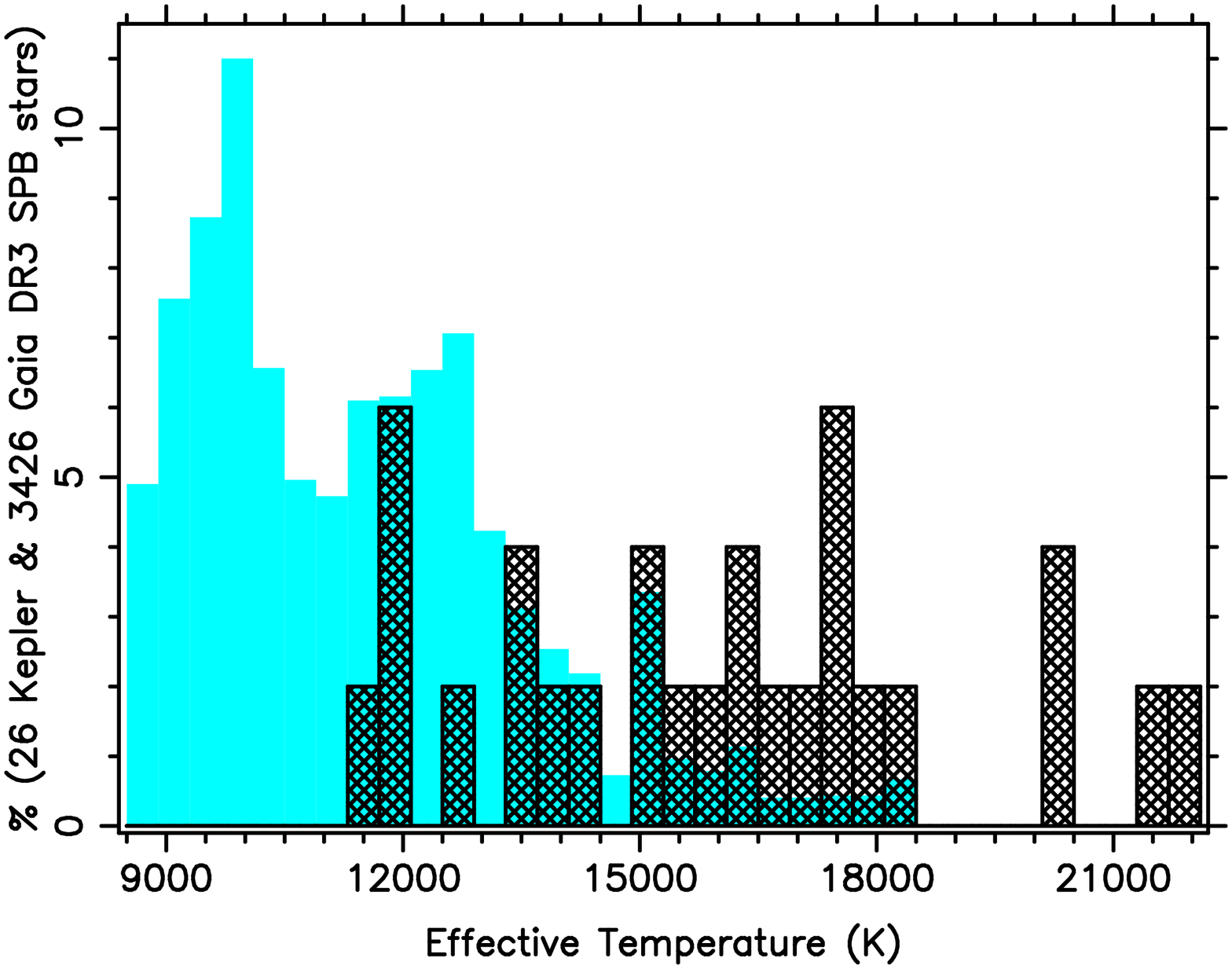}}}
\end{center}
\caption{\label{hist-teff} Histograms (normalised to 100\% occurrence)
  of the {\tt gspphot} values for $T_{\rm eff}$
  for 11636 Gaia DR3 $\gamma\,$Dor stars (left, grey) and 3426
  SPB stars (right, cyan). The width of the bars is according to the
  average error (150\,K for $\gamma\,$Dor and 400\,K for the SPB
  stars).  Asteroseismic values obtained from {\it Kepler\/} data are
  shown for 37 $\gamma\,$Dor (grey, hatched) and 26 SPB stars (black
  cross-hatched), respectively. For the right panel, {  31} SPB stars with
  a temperature above 22\,000\,K in the Gaia DR3 sample were omitted
  for visibility reasons.}
\end{figure*}
\begin{figure*}[h!]
\begin{center} 
\rotatebox{270}{\resizebox{6.1cm}{!}{\includegraphics{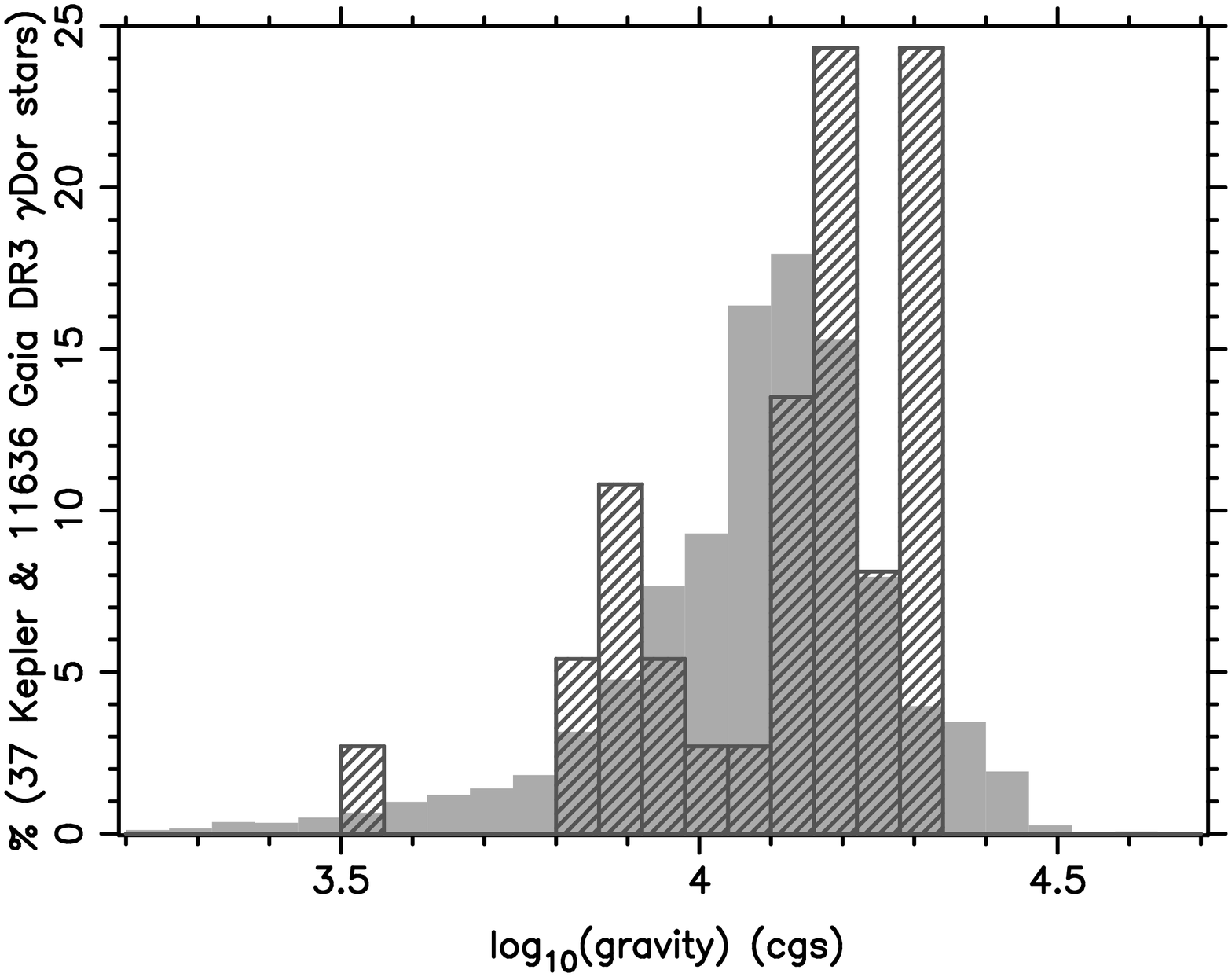}}}\hspace{0.5cm}
\rotatebox{270}{\resizebox{6.1cm}{!}{\includegraphics{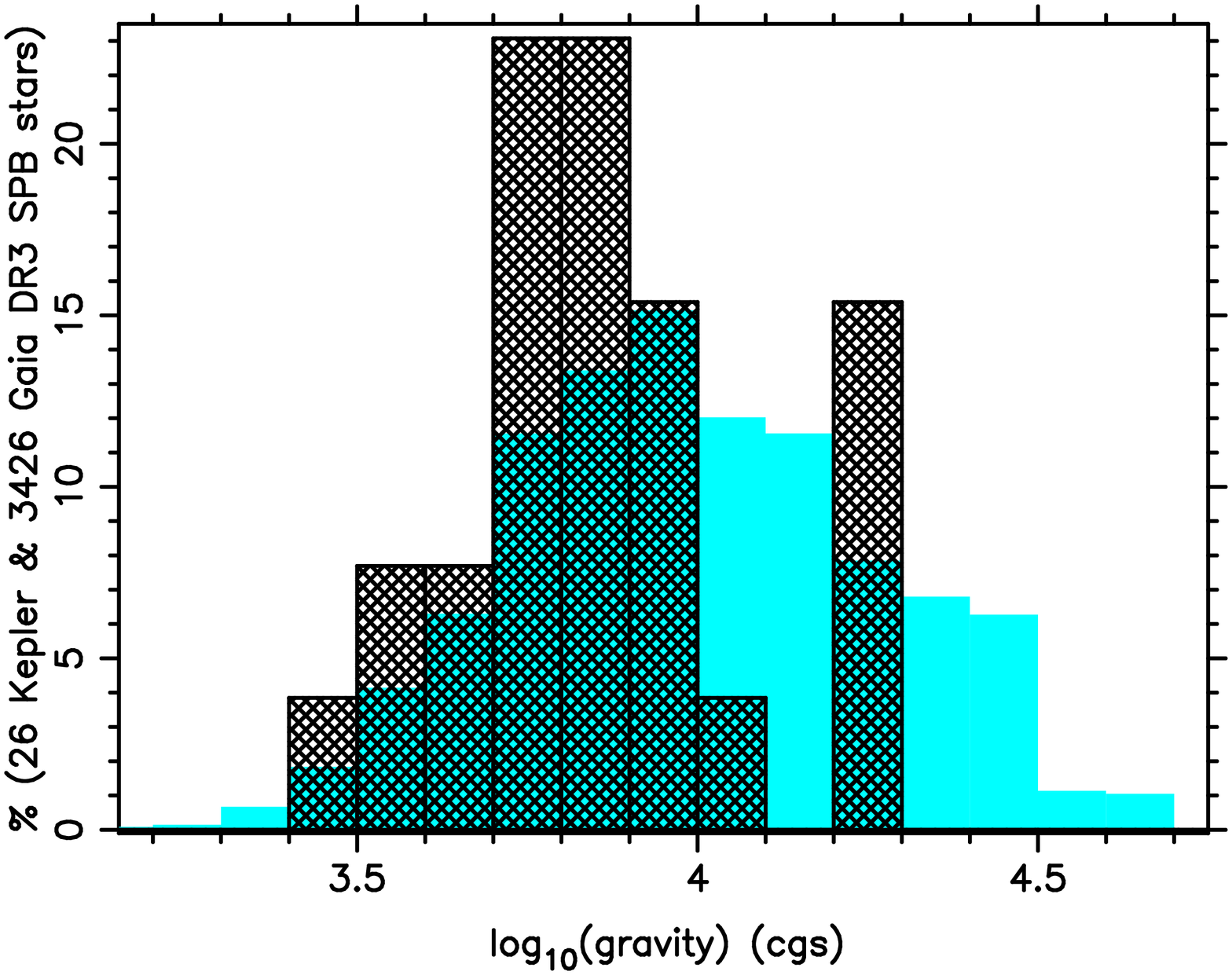}}}
\end{center}
\caption{\label{hist-logg} Same as Fig.\,\ref{hist-teff} but for $\log\,g$.}
\end{figure*}
\begin{figure*}[h!]
\begin{center} 
\rotatebox{270}{\resizebox{6.1cm}{!}{\includegraphics{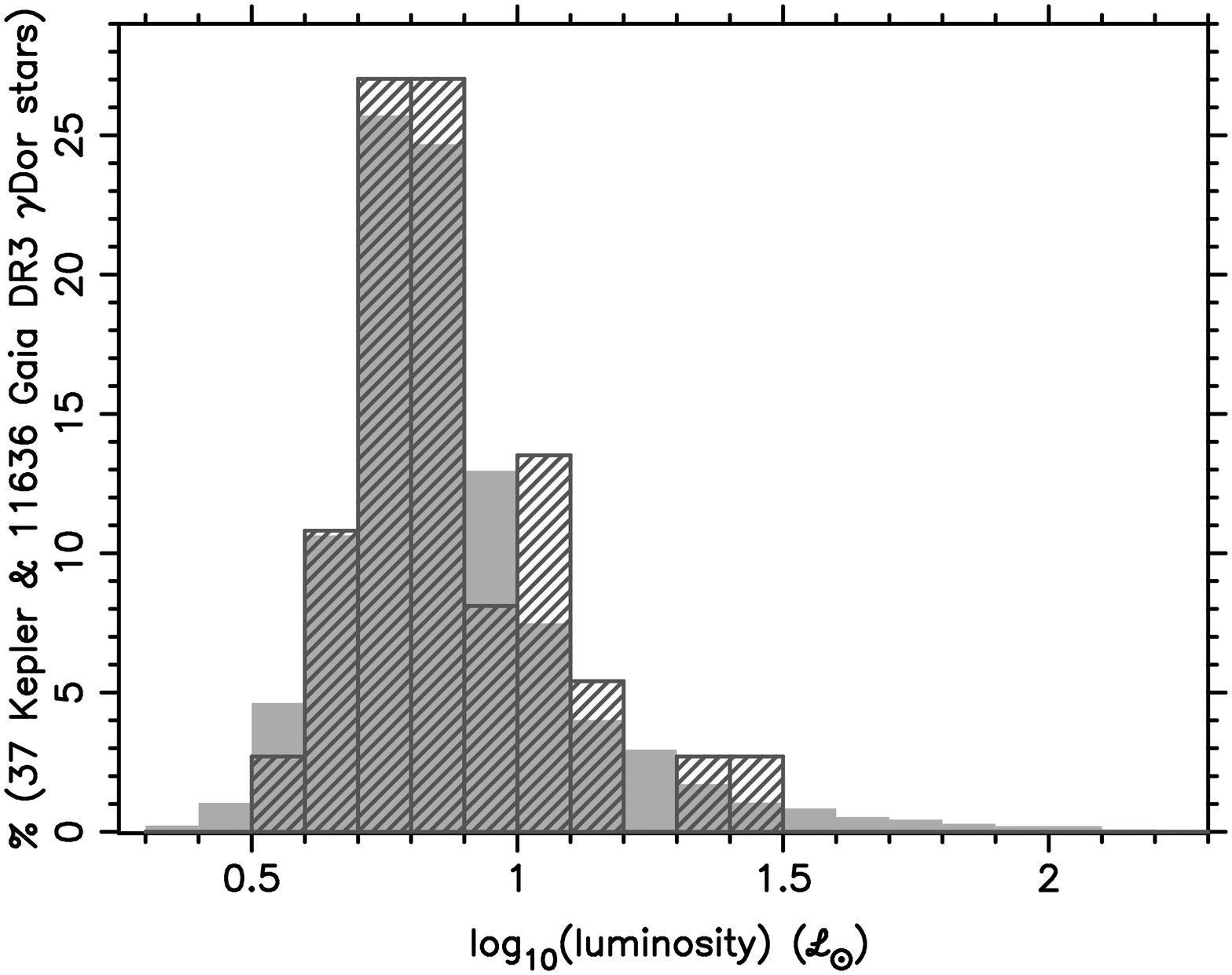}}}\hspace{0.5cm}
\rotatebox{270}{\resizebox{6.1cm}{!}{\includegraphics{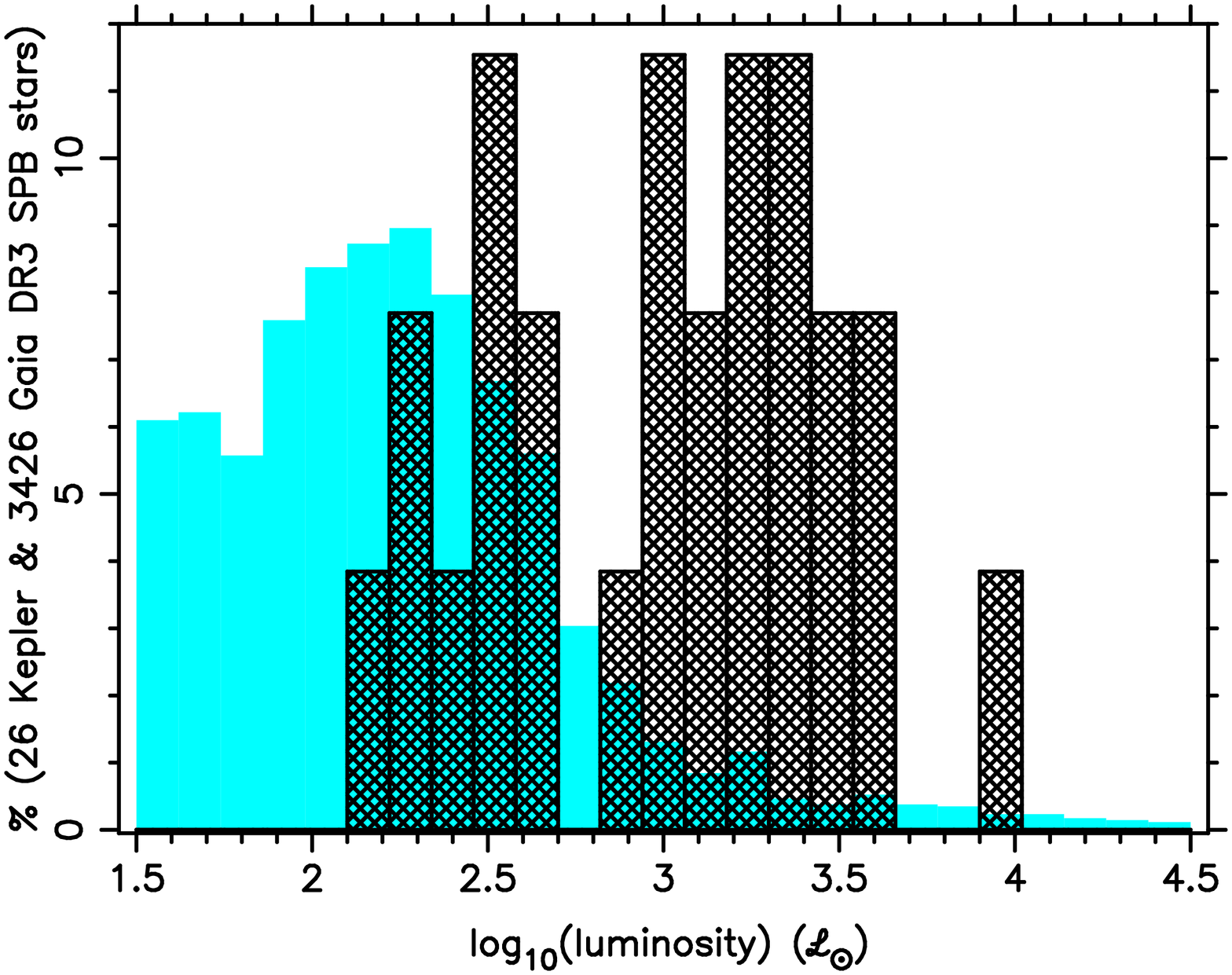}}}
\end{center}
\caption{\label{hist-lum} Same as Fig.\,\ref{hist-teff} but for $\log\,(L/{\cal L_\odot})$.}
\end{figure*}

\begin{figure*}
\begin{center} 
\rotatebox{270}{\resizebox{6.1cm}{!}{\includegraphics{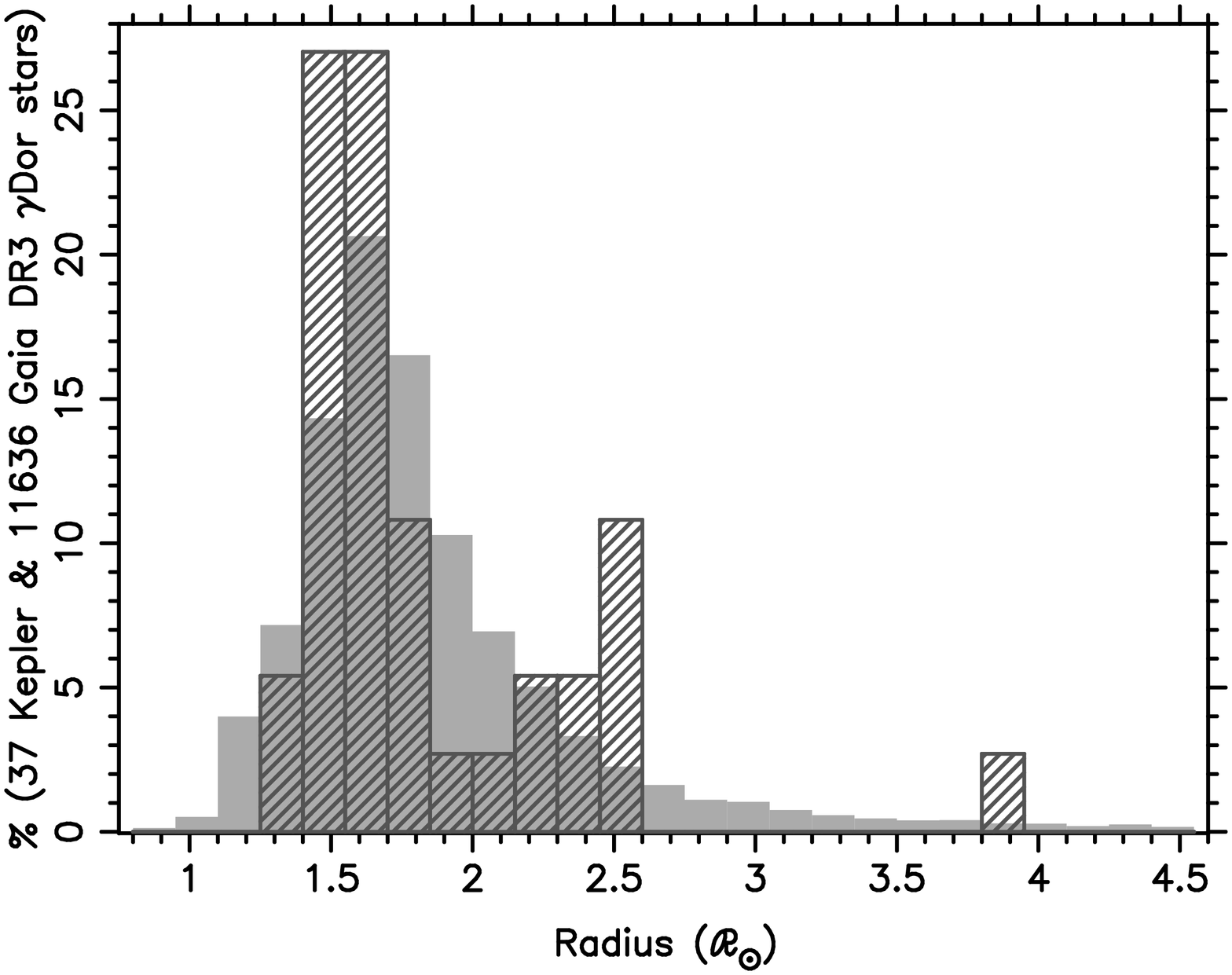}}}\hspace{0.5cm}
\rotatebox{270}{\resizebox{6.1cm}{!}{\includegraphics{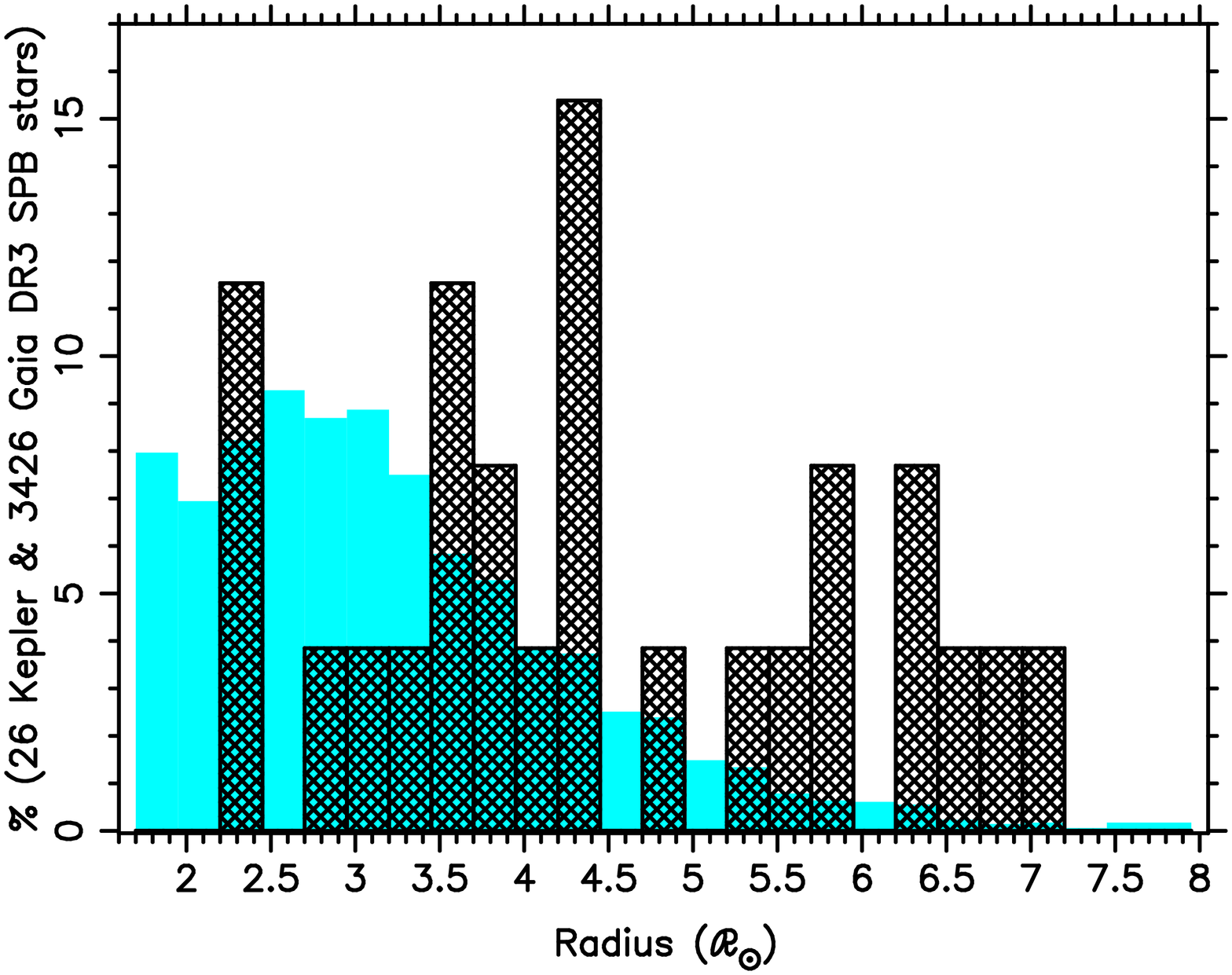}}}
\end{center}
\caption{\label{hist-radius} Same as Fig.\,\ref{hist-teff} but for the
  radius of the stars deduced from $\log\,(L/{\cal L_\odot})$ and
  $T_{\rm eff}$.}
\end{figure*}

Figures\,\ref{hist-teff}, \ref{hist-logg}, and \ref{hist-lum} show the
histograms of the effective temperature, surface gravity, and luminosity for
all 15062 Gaia DR3 g-mode pulsators in our two samples taken from the
{\tt gspphot} tables, in comparison with those quantities deduced for
the 63 bona fide pulsators.  For the latter, we took the
high-precision values for these quantities from detailed
asteroseismic modelling of their interior based on numerous identified
dipole g modes by \citet[][26 SPB stars]{Pedersen2022a} and
\citet[][37 $\gamma\,$Dor stars]{Mombarg2021}. Both these
asteroseismic studies followed the methodology in \citet{Aerts2018}
for the modelling of the internal stucture of these stars. In order to
be able to compare the distributions of the samples with vastly
different numbers of stars, we show the normalised histograms as
percentages of the entire sample population.  For the $\gamma\,$Dor
stars, the distributions of the effective temperature ($T_{\rm eff}$)
and luminosity ($\log (L/{\cal L_\odot}$), with ${\cal L}_\odot$ the
solar luminosity) agree remarkably well between the 37 bona fide
versus {  11636} Gaia DR3 pulsators, revealing Gaia's power to deduce
these two fundamental parameters for large samples of such hot F-type
pulsators. The gravities are somewhat lower than the asteroseismic
values. Comparing the Gaia radii deduced from the DR3 $T_{\rm eff}$
and $\log\,(L/{\cal L_\odot})$ values shown in the left panel of
Fig.\,\ref{hist-radius} reveals that they are entirely compatible with
the asteroseismic distribution, keeping in mind that the latter only
consists of 37 pulsators.

As for the SPB stars, the Gaia DR3 sample is concentrated in the lower
part of the SPB instability strip as already found in Paper\,I. This
paper also reported on possible systematic biases in the astrophysical
parameters of hot stars as derived from the Gaia DR3 {\tt gspphot}
tables, more particularly shifts to lower temperatures due to poorly
estimated reddening. Since we only work with the dominant frequency,
we cannot exclude that some of the (reclassified) cool SPB stars are
actually early A-type stars with rotational modulation. The majority
of {  3426} SPB stars in our sample have relatively low Gaia luminosities
compared to the 26 SPB stars in the asteroseismic sample. These two
aspects combined limit the power of Gaia DR3 to estimate radii based
on the {\tt gspphot} tables for the entire class of SPB stars, yet the
radius distribution of the {3426} SPB stars is compatible with the one
of the 26 bona fide SPB stars (Fig.\,\ref{hist-radius}, right panel).

Overall, the distributions for the three fundamental parameters
$T_{\rm eff}$, $\log\,g$, and $\log\,(L/{\cal L}_\odot)$ of the Gaia
$\gamma\,$Dor and SPB stars are in good agreement with the
asteroseismic values of the bona fide {\it Kepler\/} pulsators in
these two classes, keeping in mind that the Gaia sample of SPB stars
mainly contains cool class members.
We conclude that Gaia DR3 {\tt gspphot} values reveal
proper distributions of radii for g-mode pulsators
from their luminosity and effective temperature estimates when
compared with the radii of the 63 bona fide pulsators from the
best asteroseismic model for each of them.

\section{Pulsational properties of the dominant g modes}

\begin{figure*}[th!]
\begin{center} 
\rotatebox{270}{\resizebox{6.1cm}{!}{\includegraphics{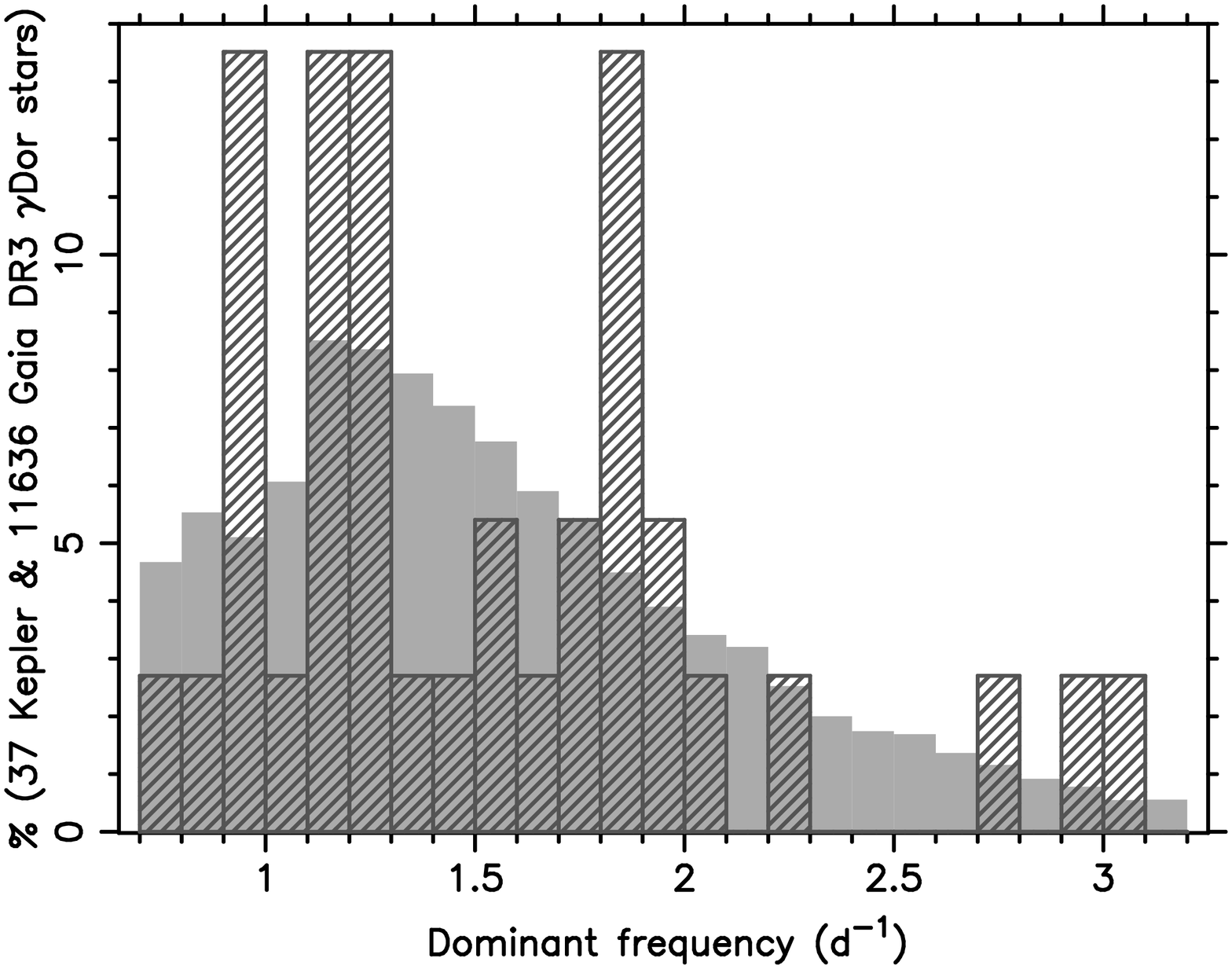}}}\hspace{0.5cm}
\rotatebox{270}{\resizebox{6.1cm}{!}{\includegraphics{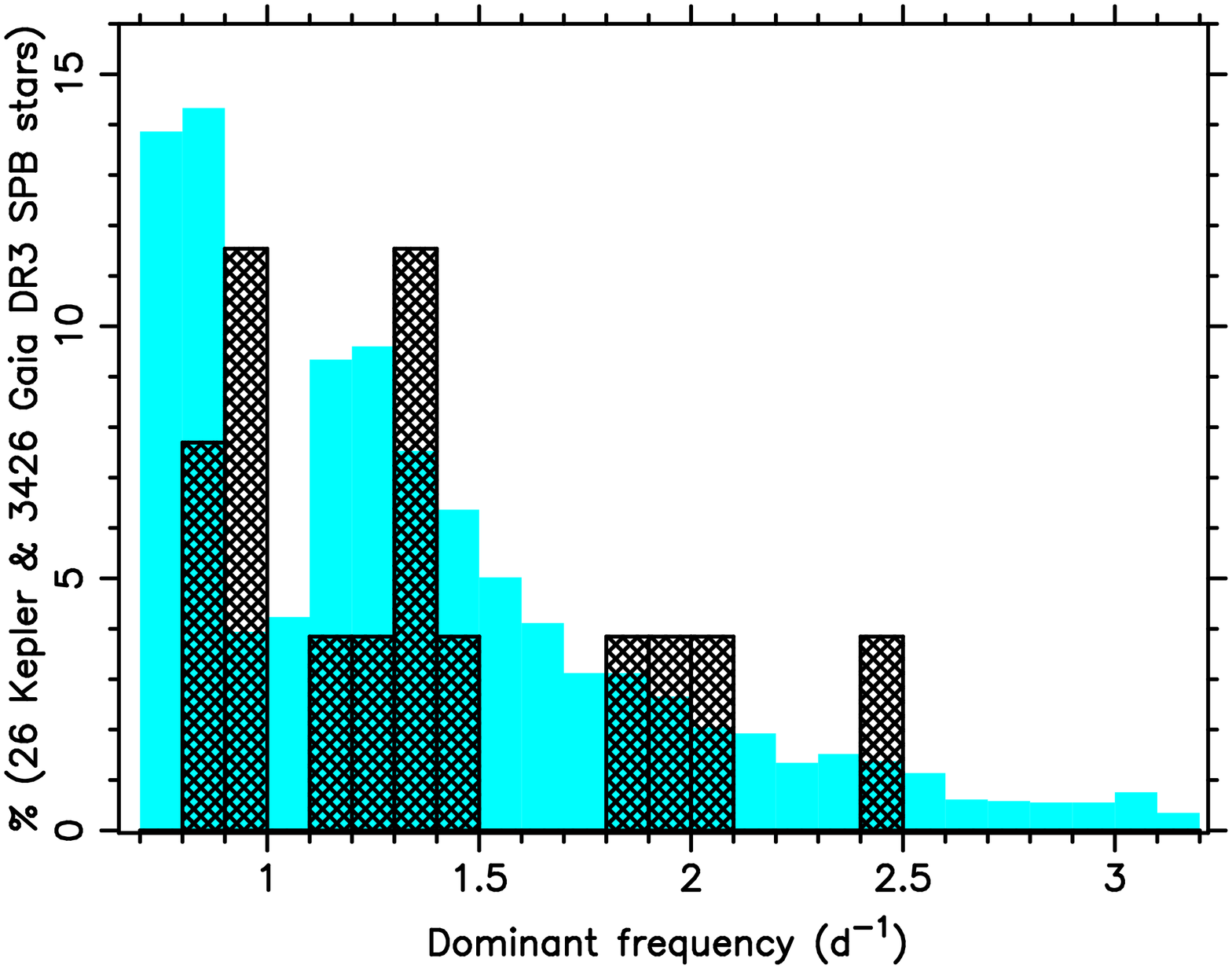}}}
\end{center}
\caption{\label{hist-freq} Same as Fig.\,\ref{hist-teff} but for the
  dominant frequency in the Gaia G light curve.}
\end{figure*}
\begin{figure*}[t!]
\begin{center} 
\rotatebox{270}{\resizebox{6.1cm}{!}{\includegraphics{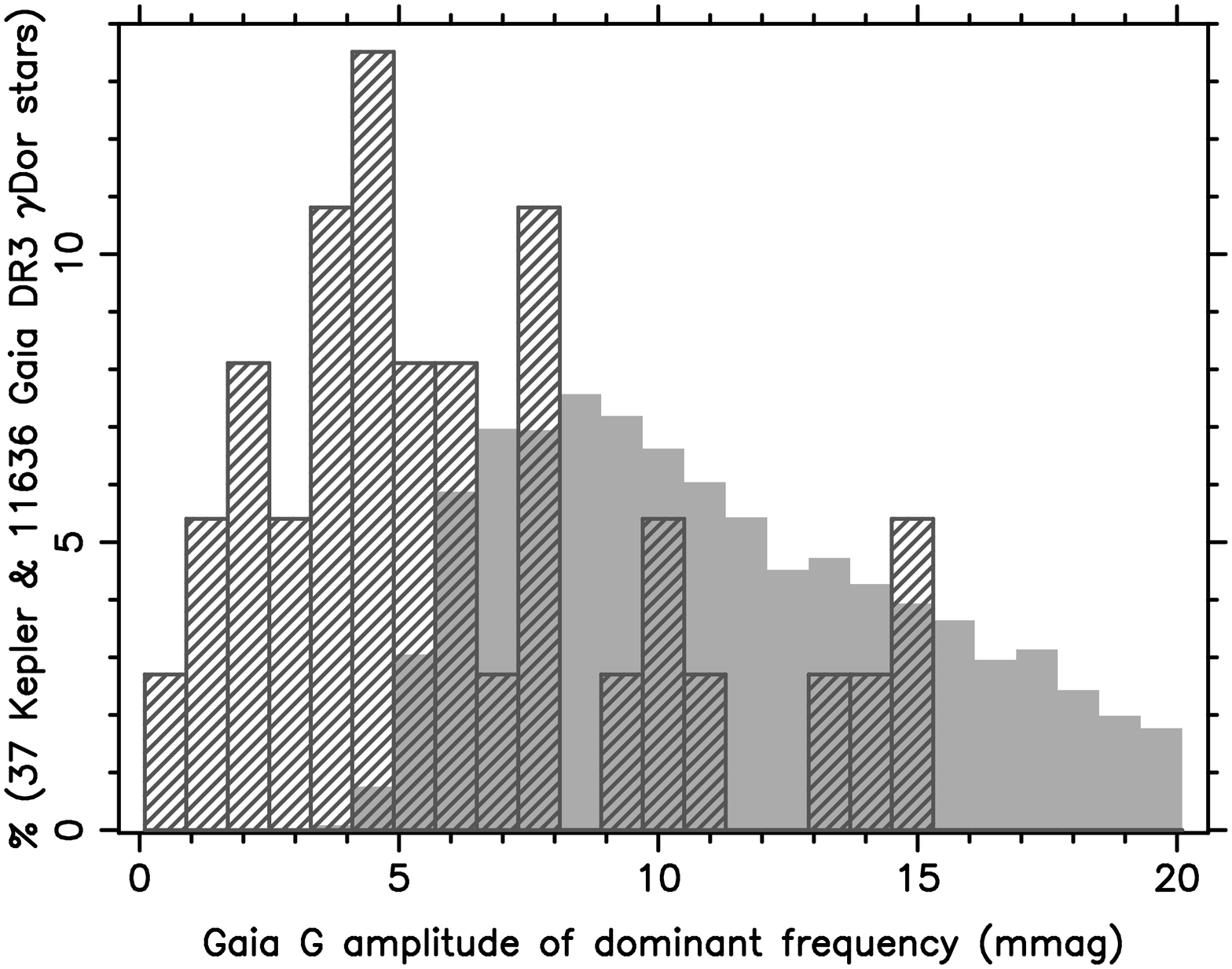}}}\hspace{0.5cm}
\rotatebox{270}{\resizebox{6.1cm}{!}{\includegraphics{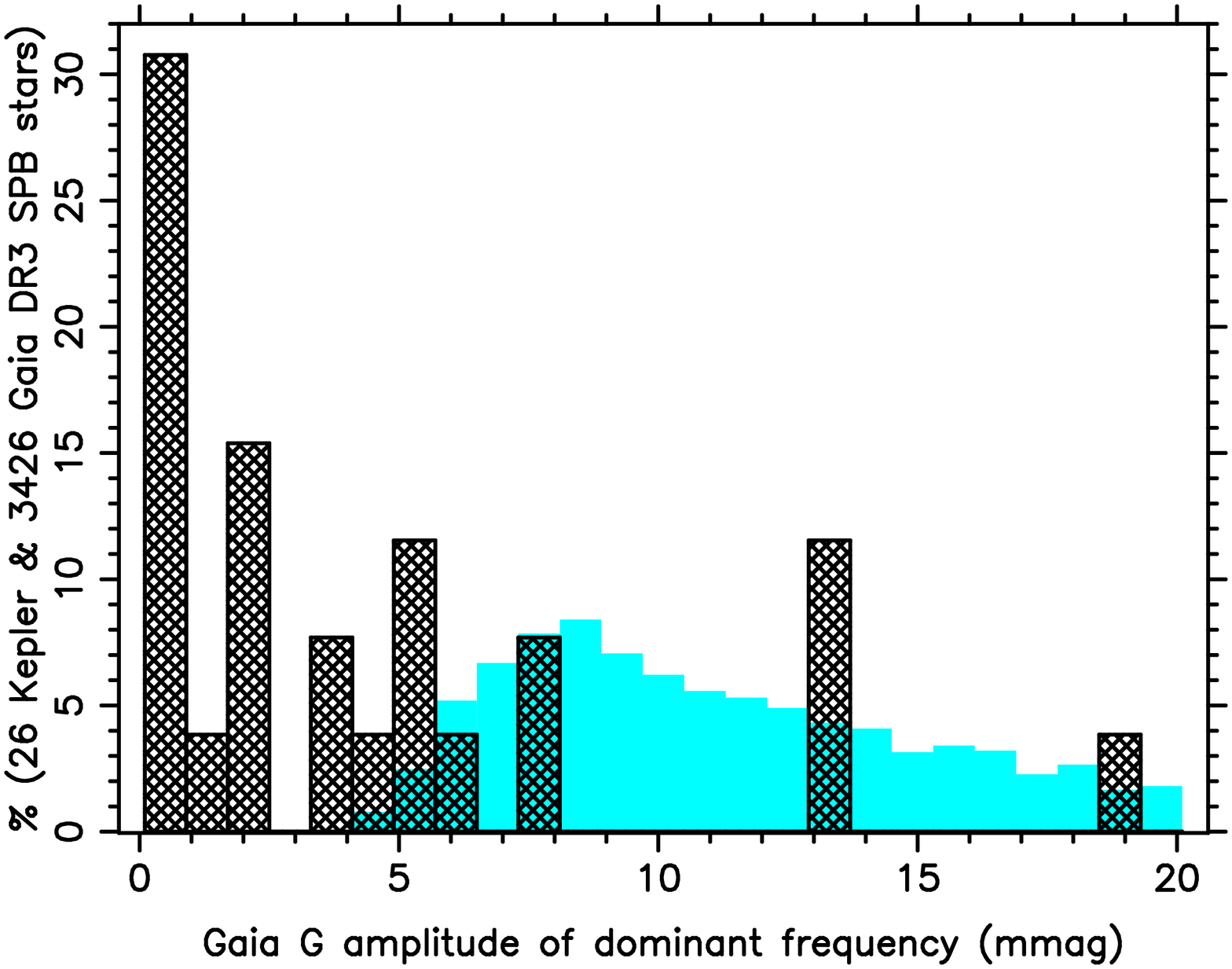}}}
\end{center}
\caption{\label{hist-ampl} Same as Fig.\,\ref{hist-teff} but for the
  amplitude of the dominant frequency in the Gaia G light curve. For
  the bona fide $\gamma\,$Dor and SPB pulsators, we computed the
  amplitude by fitting a harmonic signal to the Gaia time series using
  the main frequency derived from the 4-yr {\it Kepler\/} light
  curve.}
\end{figure*}

As already emphasised in Paper\,I, Gaia has good capacity to detect
non-radial oscillation modes in main-sequence stars. The two Gaia g-mode
pulsator samples treated here result from quite strict selection
rules on the Gaia G photometric light curves,
yet they are already an order of magnitude larger than the
corresponding {\it Kepler\/} samples. Despite the sparse Gaia
sampling, it is to be anticipated that many more g-mode pulsators will
be selected once the DR4 and DR5 data sets will become available.

Figure\,\ref{hist-freq} shows the distributions for the dominant
frequency in the DR3 Gaia G band light curves. Overplotted are the
distributions for the dominant frequency deduced from the far
higher-precision and 4-yr uninterrupted high-cadence {\it Kepler\/}
light curves taken from \citet{VanReeth2015} and \citet{Pedersen2021}
for the $\gamma\,$Dor and SPB stars, respectively. We recall that we used
the dominant g-mode frequency range covered by these two samples of
bona fide g-mode pulsators as selection criterion, to
restrict the Gaia DR3 samples to pulsators adhering to this same
appropriate frequency range. Hence it is built in that we find
compatible ranges. Yet, also the distributions are reasonably well in
agreement between the Gaia DR3 and {\it Kepler\/} pulsators, keeping
in mind the small samples sizes for the latter.

Figure\,\ref{hist-ampl} shows the histograms for the amplitude of the
dominant frequency found in the Gaia G light curve. These bona fide
pulsators did not survive our six selection criteria, mainly because
they have fewer than 40 epochs in Gaia DR3 and/or their dominant
frequency did not meet the FAP criterion.  In order to be able to
compare the amplitudes between the Gaia DR3 and {\it Kepler\/}
pulsators, and to exclude instrumental effects for the bona fide
pulsators, we computed their Gaia G amplitude by imposing their
dominant frequency found in their {\it Kepler\/} light curve onto the
Gaia G data, irrespective of the number of epochs in the latter and
their FAP value.  Both histograms in Fig.\,\ref{hist-ampl} visualise
the current detection threshold to find g modes in the Gaia G light
curves. It is seen that DR3 allows us to detect g-mode frequencies
with an amplitude above 4\,mmag. The {\it Kepler\/} data delivered g
modes with far lower amplitudes as seen in the histogram, because the
mission was designed to assemble $\mu$mag-precision uninterrupted
30-min cadence photometry for exoplanet hunting \citep{Borucki2010}
and for asteroseismology \citep{Gilliland2010}.  We find that the
g-mode amplitude distributions for Gaia DR3 and {\it Kepler\/} hardly
overlap for the class of the SPB pulsators as most of the 26 bona fide
{\it Kepler\/} SPB stars have low dominant amplitudes outside Gaia's
reach.

\section{Properties of spectral line broadening}

Aside from photometric and astrometric data, the Gaia satellite also
delivers spectroscopic data. Its spectrometer RVS has a median
resolving power of 11\,500 and is sensitive to the wavelength range from
846 to 870 nm. While it was built with the primary goal to measure the
radial velocity of as many Gaia sources as possible \citep{Katz2022},
we use the RVS data to study the spectral line broadening of
g-mode pulsators. Our aim is to investigate if, aside from
rotational line broadening, there is any connection between the
overall line broadening, the fundamental parameters, and the
oscillation properties for the two large Gaia DR3 samples of g-mode
pulsators, as suggested previously based on line-profile simulations
\citep{Aerts2009,AertsRogers2015}.  While RVS on average provides a
resolving power of only $\sim\! 26\,$km\,s$^{-1}$, non-radial
oscillations generate variations in the width and the skewness of
spectral lines \citep{AertsDeCat2003} and these may affect
the way that the line broadening values were determined \citep[we
  refer to][for a detailed description omitted here]{Fremat2022}.

Line-profile variations caused by the g~modes of $\gamma\,$Dor and SPB
stars occur at the level of several to tens of km\,s$^{-1}$ in the
centroid of the line
\citep[e.g.,][]{Aerts1999,DeCat2000,Mathias2001,DeCatAerts2002,Mathias2004,DeCat2006}.
High-resolution time-series spectroscopy of bright g-mode pulsators
offers a powerful tool to identify the spherical wavenumbers $(l,m)$
of the dominant oscillation mode(s) provided that the oscillation
cycle is well covered \citep{Briquet2003,DeCat2005}. Such applications
couple the velocity field, computed from the theory of non-radial
oscillations, to the observed line-profile variations to infer the
radial and tangential components of the velocity vector due to each
non-radial oscillation mode \citep[][Chapter\,6]{Aerts2010}.  This
requires the spectroscopy to have high resolving power and
signal-to-noise ratio (SNR) \citep[typically above 50\,000 and 300,
  respectively;][]{AertsDeCat2003}.

In the absence of high-quality spectroscopy, or in the case where only
a few snapshot spectra are available, line-profile modelling by means
of the proper time-dependent pulsational velocity field is
impossible. In such a case, it is customary to approximate the overall
line broadening due to oscillations and rotation together by a single
time-independent function called macroturbulence \citep{SSD2010}. Even
though its functional form assumes a symmetrical line profile
\citep[cf.\,][Fig.\,9]{Aerts2014b}, the macroturbulence correlates
strongly with quantities representing the line-profile variability
\citep{SSD2017}, such as the velocity moments
\citep{Balona1986,Aerts1992}.

Fitting time-resolved line-profile variations due to oscillations or
spots artificially with a symmetrical macroturbulent profile leads to
time-variability in the macroturbulence which is in excellent
agreement with the mode frequencies or rotation periods of
intermediate-mass dwarfs \citep{Aerts2014b}.  This suggests that
macroturbulence is merely a downgraded (and often poor)
time-independent symmetrical simplification of the true spectral line
profiles caused by oscillations and/or spots.  Nevertheless, in the
absence of time-resolved spectroscopy, it is a sensible approach to
fit the line profiles of snapshot spectra with a synthetic
time-independent macroturbulent broadening profile, particularly for
large surveys of stars such as offered by Gaia.

The g~modes have dominant tangential displacements, implying that
their velocity at the limb of the star dominates the detected
line-profile variability.  Yet, a common attitude in the literature
has been to rely on the ad-hoc assumption that the radial and
tangential components of the macroturbulent broadening profile are
equal \citep{SSD2014}. This is the reason why unrealistic, often
supersonic, values for the macroturbulent surface velocities are
obtained. This in turn affects the estimation of the surface rotation
\citep{Aerts2014b}. For that reason, it is essential to estimate the
surface rotation first, independently from the macroturbulent
broadening \citep{Nadya2022}.

In the following sections we investigate Gaia's capacity to shed light
on the astrophysical cause(s) of its measurements of the spectral line
broadening from RVS. We do so for the two classes of $\gamma\,$Dor and
SPB stars, whose velocity field at the stellar surface due to their
non-radial oscillations is dominantly tangential
\citep{DeCatAerts2002,Aerts2004}.

\subsection{Gaia DR3 line broadening parameters}

While Gaia's medium-resolution RVS was not built to assess line
broadening by stellar oscillations combined with rotation, it offers
unprecedently large stellar samples analysed with a common
methodology. We test the behaviour of line broadening measured with
RVS with respect to the acting velocity fields at the stellar surface,
where we know that our two samples undergo the joint effect of
time-independent rotational and time-dependent pulsational line
broadening.  To do so, we rely on two Gaia DR3 parameters offered as
measurement of spectral line broadening: {\tt vbroad}
\citep{Fremat2022} and {\tt vsini$_{-}$esphs} \citep{Creevey2022}.
\citet{Fremat2022} already made a careful study of these two
quantities to interpret spectral line broadening for more than 33
million stars having $T_{\rm eff} \in [3.1;14.5]$kK. This range fully
encapsulates the one of our $\gamma\,$Dor sample and largely overlaps
with the SPB sample.

\begin{figure}[th!]
\begin{center} 
\rotatebox{270}{\resizebox{7.5cm}{!}{\includegraphics{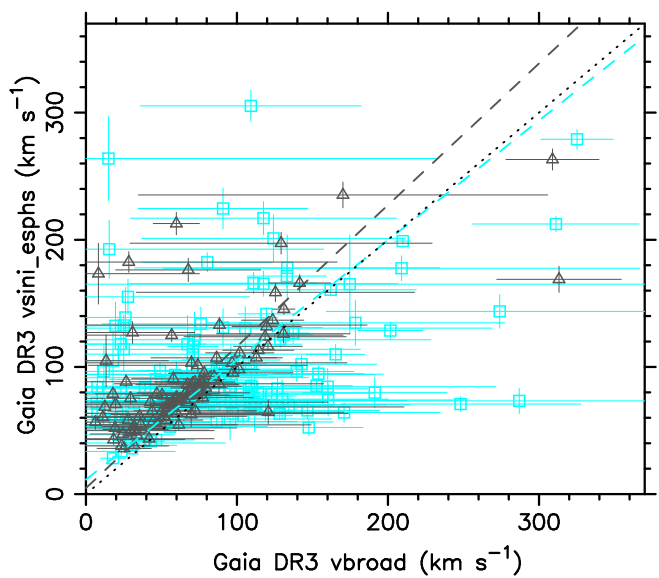}}}
\end{center}
\caption{\label{vbroad-vsini} Comparison between Gaia DR3 measurements
  of {\tt vsini$_{-}$esphs} and {\tt vbroad} for the {  100} $\gamma\,$Dor
  (grey triangles) and {  156} SPB (cyan squares) stars having both
  quantities available. The full dotted line indicates the bisector
  while the coloured dashed lines represent the best linear regression
  models for both samples.}
\end{figure}

The Gaia DR3 parameter {\tt vbroad} captures the overall line
broadening after deconvolving the spectra with Gaia's RVS instrument
Along-Scan Line Spread Function \citep{Sartoretti2022}. This parameter
{\tt vbroad} thus includes the joint effect of all possible
astrophysical causes that give rise to spectral line broadening, such
as oscillations, rotation, spots, turbulent convective velocities,
etc.  The catalogued value of {\tt vbroad} is the median value
obtained by the Multiple Transit Analysis (MTA) over at least six
valid transits.  The corresponding catalogued uncertainty is the
standard deviation with respect to this median value. This implies
that the uncertainty may be a measure of the line-profile variability,
because its range captures the line broadening occurring at minimally
six different epochs aside from the contribution of the noise.
For the current study, we have {\tt vbroad}
measurements for {  1775} of the {  11636} $\gamma\,$Dor stars and
for {  190} of the {  3426} SPB pulsators.

Another estimate of the RVS line broadening denoted as {\tt
  vsini$_{-}$esphs} is obtained by the so-called Extended Stellar
Parametrizer module developed within the Astrophysical ParameterS
Inferences System APSIS \citep{Creevey2022}. APSIS is able to treat
the parameters of hot stars and delivers {\tt vsini$_{-}$esphs} as an
intermediate data product approximating time-independent rotational
broadening. Its computation was done on the basis of the averaged
values of BP and RP and the averaged RVS spectrum.
Although this implies the
variability to be filtered out from the quantity {\tt vsini$_{-}$esphs}
to some level, it still contains a contribution from the oscillations
\citep[cf.,][for the theoretical expression of the line width due to
  non-radial oscillations]{DePauw1993}. 
Nevertheless, {\tt vsini$_{-}$esphs}
is a cleaner measurement of the time-independent
projected surface rotation velocity
than  {\tt vbroad}
when 
rotation dominates the spectral line broadening, as is the case for
g-mode pulsators. The error published
for {\tt vsini$_{-}$esphs} is an approximation of the statistical
error and does not represent a measurement of the time-variable line
broadening like the standard deviation for {\tt vbroad}. The quantity
{\tt vsini$_{-}$esphs} deduced by APSIS results from an optimal RVS
and BP/RP treatment for stars having a $T_{\rm eff}$ above 7500\,K,
while it relies on BP/RP alone for all stars with $T_{\rm eff} >
7000\,$K that were not observed by RVS.  Values for {\tt
  vsini$_{-}$esphs} are available for {  384} of the {  11636} $\gamma\,$Dor
stars and for {  1104} of the {  3426} SPB stars.

The wavelength coverage of RVS was constructed so as to achieve
optimal radial-velocity data for a broad range of stellar
populations. Here we rely on its data for the purpose of studying
spectral line broadening, for which the RVS wavelength domain is not
optimal. This is particularly so for the hottest stars under study
here.  As a consequence, some of the {\tt vbroad} and {\tt
  vsini$_{-}$esphs} measurements have large uncertainties. It is
therefore essential to include these uncertainties, aside from the
values themselves, in any proper astrophysical interpretation. We
refer to \citet{Fremat2022} for a detailed and nuanced discussion
about the quality of the {\tt vbroad} and {\tt vsini$_{-}$esphs}
measurements deduced from template spectra relying on the Gaia $T_{\rm
  eff}$ estimates. In particular, \citet{Fremat2022} discussed in
detail the correlations between these two parameters for stars
covering a broad range of magnitudes and temperatures.
Of particular relevance for the current work is Fig.\,16 in
\citet{Fremat2022}, illustrating two Hertzsprung-Russell diagrams with
density maps as a function of median {\tt vbroad} values.  That figure
clearly reveals high {\tt vbroad} values along the upper main
sequence, where the $\gamma\,$Dor and SPB pulsators are situated
(encompassing the p-mode dominated class of $\delta\,$Sct stars not
treated here due to the much higher risk that their dominant frequency
has an instrumental origin, as explained in Sect.\,2). The figure
reveals higher {\tt vbroad} values for hotter stars but
\citet{Fremat2022} do not provide any formal quantitative comparisons
between {\tt vbroad} and the stellar parameters.

We show the {  100} $\gamma\,$Dor and {  156} SPB pulsators having a
measurement of both {\tt vbroad} and {\tt vsini$_{-}$esphs} in
Fig.\,\ref{vbroad-vsini}, along with their uncertainties. It can be
seen that the overall range of the two quantities is rougly the same
for these $\gamma\,$Dor and SPB stars. For each of the stars in
Fig.\,\ref{vbroad-vsini}, the two plotted quantities have similar yet
not always equal values according to the uncertainty estimates.
We remind that similarities between the
two samples as a whole also occur for their ranges of the dominant
mode amplitude and mode frequency (cf.\,Figs\,\ref{hist-freq} and
\ref{hist-ampl}). It is then a natural question whether the oscillation
properties cause the time-variability measured by {\tt vbroad} and its
standard deviation. On the other hand, we investigate whether the
decrease of observed mode amplitude for faster rotators, as found in
Paper\,I for the dominant p~modes of the Gaia DR3 $\delta\,$Sct stars,
also occurs for g-mode pulsators.

In what follows we offer regression models accomodating 
errors-in-variables, which allows one to handle different
measurements of the same astrophysical quantity (here the overall
time-averaged spectral line broadening) having both star-specific and
measurement-specific errors. These errors must be propagated properly
when constructing the regression models and interpreting their
outcome. This has been used in the context of Gaia data before, for example
in a comparison between asteroseismic and astrometric parallaxes
following DR1 \citep{DeRidder2016}. We first provide a general
description of the methodology. Subsequently we apply it to the sample
of the 37 bona fide $\gamma\,$Dor stars. For all these 37 stars we
also have, in addition to their Gaia DR3 data, estimates of their
``true'' pulsational and rotational line broadening deduced from one
homogeneous treatment of high-resolution high signal-to-noise
ground-based spectroscopy taken with one instrument (such homogeneous
spectroscopic information is not available for all 26 bona fide
SPB pulsators).  We use the results obtained for the 37 bona fide
$\gamma\,$Dor stars to treat the Gaia DR3 g-mode samples optimally
with the aim to interpret their spectral line broadening properties.

\subsection{Errors-in-variables model}

Let us denote two observed quantities by $Y_i$ and $X_i$ and their true but
unknown values by $Y_i^\ast$ and $X_i^\ast$.
The errors-in-variables model is then specified by:
\begin{eqnarray}
Y_i&=&Y_i^\ast+\varepsilon_{Yi},\label{een}\\
X_i&=&X_i^\ast+\varepsilon_{Xi},\label{twee}\\
Y_i^\ast&=&\beta_0+\beta_1 X_i^\ast+\varepsilon_i,\label{drie}
\end{eqnarray}
with $i$ indexing the stars in a sample.
Here, $\beta_0$ and $\beta_1$ are fixed but unknown regression
coefficients to be estimated from the data.
The measurement error variances $\mbox{var}(\varepsilon_{Yi})=\sigma^2_{Yi}$ and 
$\mbox{var}(\varepsilon_{Xi})=\sigma^2_{Xi}$ are
obtained from the observations.
The residual error component $\varepsilon_i$, with variance
$\sigma^2$,
quantifies imperfection in the regression relationship. 

If $Y_i^\ast$ and $X_i^\ast$ would be nearly identical,
then $\beta_0$ is expected to be close to 0 and $\beta_1$
would be close to 1.
If the regression relationship in Eq.\,(\ref{drie}) is very precise
relative
to the measurement error, then $\sigma^2$ would be near~0. 
Expressions\,(\ref{een})--(\ref{drie}) yield the mean and variance relationships:
\begin{eqnarray}
E(Y_i)&=&\beta_0+\beta_1 X_i,\\
\mbox{Var}(Y_i)&=&\beta_1^2\sigma^2_{Xi}+\sigma^2_{Yi}+\sigma^2.
\end{eqnarray}
Assuming (approximate) normality, a fully parametric specification follows, thus
enabling
maximum likelihood estimation:
\begin{equation}\label{vijf}
Y_i\sim N(\beta_0+\beta_1 X_i,\beta_1^2\sigma^2_{Xi}+\sigma^2_{Yi}+\sigma^2).
\end{equation}
The SAS procedure {\tt NLMIXED} \citep{SAS} was used
for the maximum likelihood estimation.

Extension to multiple predictors $X_{1i},\dots,X_{pi}$ is
straightforward,
upon replacing Eq.\,(\ref{vijf})  by: 
\begin{equation}\label{zes}
  Y_i\sim N\left(\beta_0+\sum_{j=1}^p\beta_j
  X_{ji},\sum_{j=1}^p\beta_j^2\sigma^2_{Xji}+\sigma^2_{Yi}+\sigma^2\right),
\end{equation}
with obvious notation. 
It is convenient to write Eq.\,(\ref{zes}) in vector notation as:
\begin{equation}\label{zesbis}
  Y_i\sim N(\mbox{\boldmath $X$}_i'\,\mbox{\boldmath $\beta$},
  \mbox{\boldmath $\beta$}'\,\Sigma_{x,i}\,\mbox{\boldmath $\beta$}+\sigma^2_{Yi}+\sigma^2),
\end{equation}
where $\mbox{\boldmath $\beta$}=(\beta_0,\beta_1,\dots,\beta_p)'$
and $\Sigma_{x,i}$ is a diagonal matrix with
$(0,\sigma^2_{X1i},\dots,\sigma^2_{Xpi})'$
along the diagonal.

Model-based prediction of $Y_i^\ast$ and its standard deviation can be expressed as:
\begin{eqnarray}
\label{zeven}
\widehat{Y}_i^\ast&=&\widehat{\beta}_0+\sum_{j=1}^p\widehat{\beta}_j X_{ji},\\
\widehat{\mbox{s.d.}}(\widehat{Y}^\ast_i)&=&\sqrt{
  \widehat{\mbox{\boldmath $\beta$}}'\Sigma_{x,i}\widehat{\mbox{\boldmath $\beta$}} +
\mbox{\boldmath $x$}_i'\widehat{\mbox{var}}(\widehat{\mbox{\boldmath
    $\beta$}})
\mbox{\boldmath $x$}_i
+\sigma^2_{Yi}+\widehat{\sigma}^2},
\label{acht}
\end{eqnarray}
where the unknown parameters have been replaced by their data-based
estimates.  Expressions\,(\ref{zeven})--(\ref{acht}) can be used to
assess the quality of the model fit.  The second term under the square
root in Eq.\,(\ref{acht}) takes the uncertainty in the estimated
regression coefficient into account.

\subsection{Spectral line broadening for the 37 bona fide $\gamma\,$Dor stars}

\begin{figure}[h!]
  \begin{center}
    \rotatebox{270}{\resizebox{6.6cm}{!}{\includegraphics{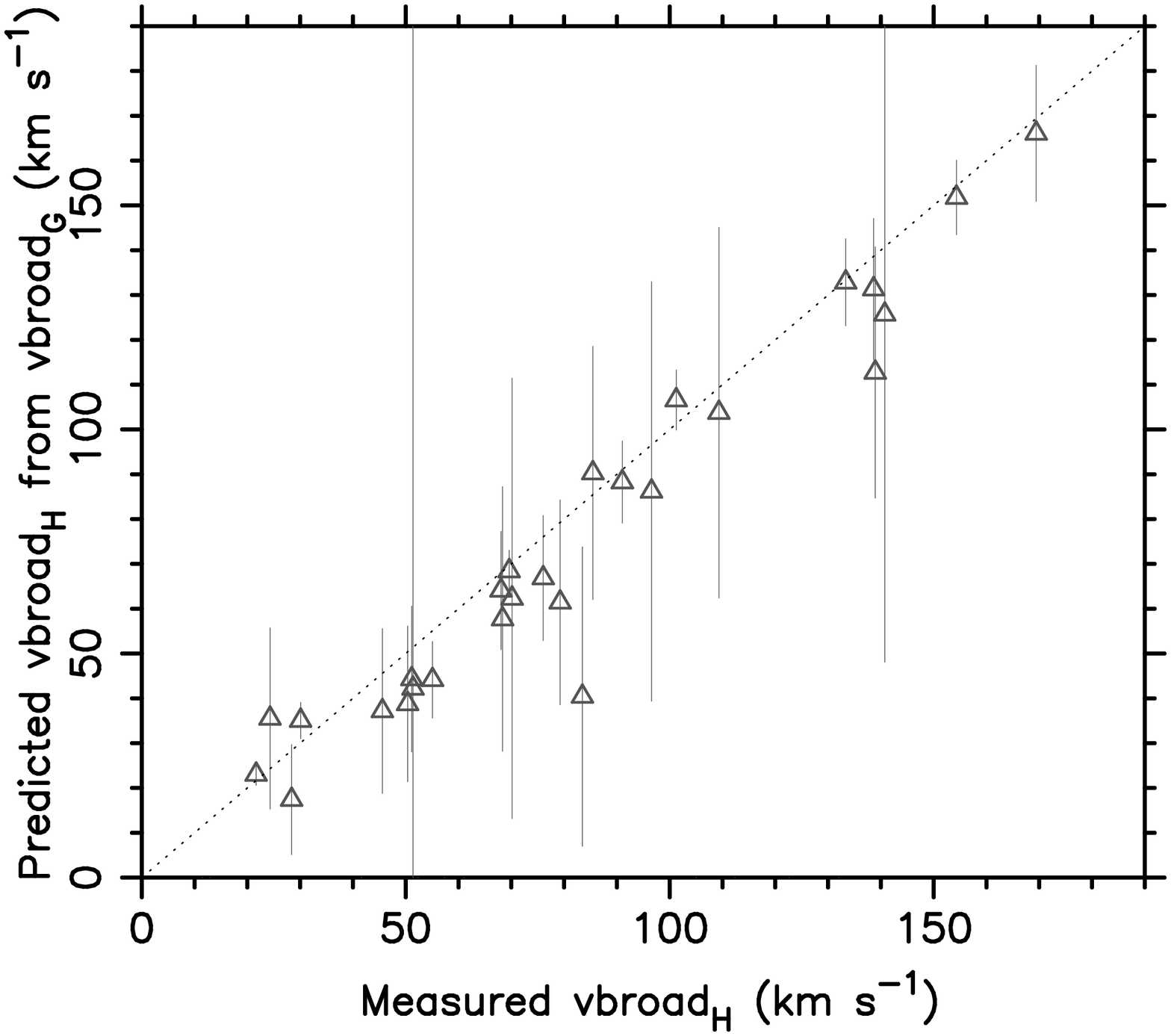}}}\vspace{0.5cm}
    \rotatebox{270}{\resizebox{6.6cm}{!}{\includegraphics{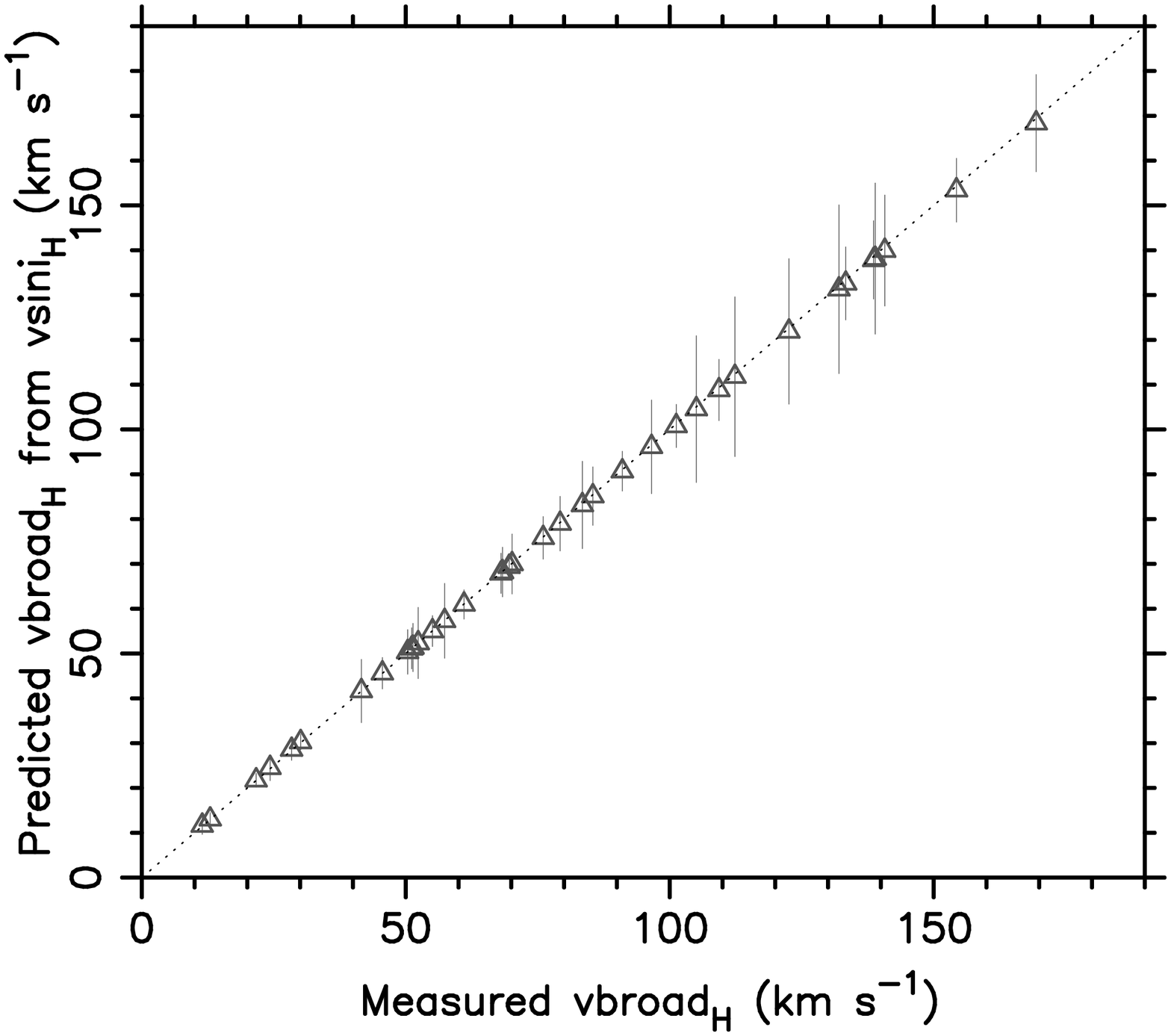}}}\vspace{0.5cm}
    \rotatebox{270}{\resizebox{6.6cm}{!}{\includegraphics{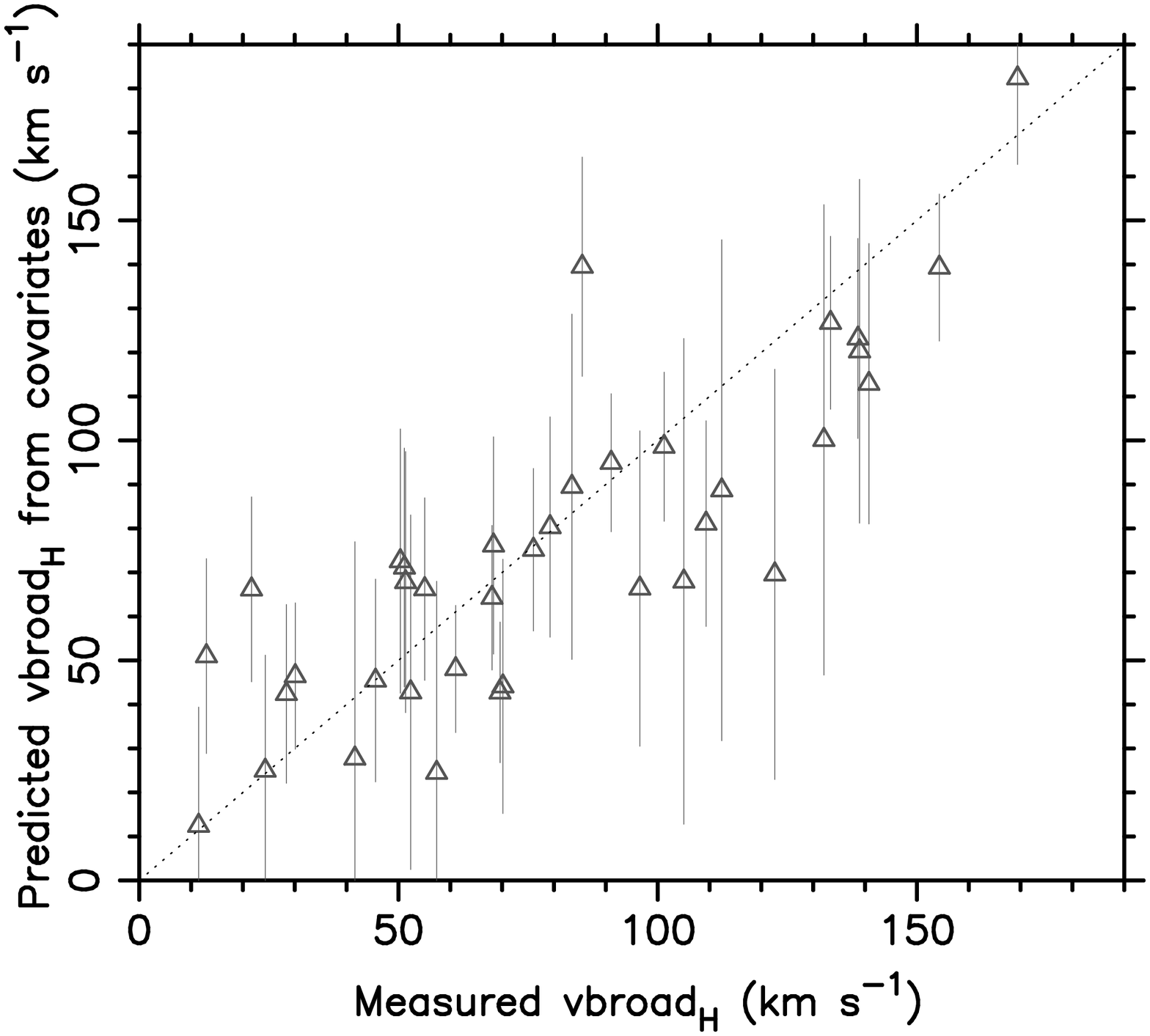}}}
\caption{Quality of predictions of {\tt vbroad$_{\rm H}$}, by {\tt
    vbroad$_{\rm G}$} (upper panel), {\tt vsini$_{\rm H}$} (middle panel), and the set of
  covariates (lower panel) for the 37 bona fide {\it Kepler\/}
  $\gamma\,$Dor stars.
  The vertical bars are defined by the
  predicted value $\pm$ its standard deviation, based on
  the errors-in-variables models in Table\,\ref{table2}.}
\label{Geert}
\end{center}
\end{figure}

For the 37 bona fide $\gamma$\,Dor stars, we now add three more
quantities to the dominant frequency $\nu$ from {\it Kepler\/}
photometry and the Gaia DR3 values for $\log\,T_{\rm eff}$, $\log\,g$,
$\log\,(L/{\cal L}_\odot$), and $A_\nu$.
Following their discovery as g-mode pulsators in the {\it
  Kepler\/} data, \citet{Tkachenko2013} set up a ground-based
spectroscopic campaign with the HERMES spectrograph attached to the
1.2m Mercator telescope situated at La Palma Observatory, Spain
\citep{Raskin2011}. HERMES has a spectral resolution of 85000 and
covers wavelengths from 377 to 900\,nm.  The HERMES spectra allowed
\citet{VanReeth2015} to deduce the overall spectral line broadening,
here denoted as {\tt vbroad$_{\rm H}$}, for the 37 stars and to
unravel it into separate components stemming from time-independent
rotational broadening ({\tt $v\sin\,i_{\rm H}$}) and a broadening
component due to the joint effect of microturbulence and the
oscillation modes at the particular
epoch of the observed spectrum. Microturbulence represents an
artificial Gaussian line broadening needed
to bring observed spectral lines into agreement with line predictions from
1-dimensional atmospheric models. This small broadening is needed to take into account 
the occurrence of small-scale turbulent motions in the line forming
region that are not included in atmosphere models. On the other hand,
the velocities due to non-radial g-mode oscillations result in time-dependent line
broadening. In the case of $\gamma\,$Dor stars, their joint net effect
in the line-of-sight is of similar order of a few km\,s$^{-1}$
\citep{Aerts2004,DeCat2006}. We therefore take the Gaussian line
broadening determined by \citet{VanReeth2015} as a good approximation
for the overall pulsational broadening and denote it as 
{\tt vosc$_{\rm H}$}.

The two quantities {\tt $v\sin\,i_{\rm H}$} and {\tt vosc$_{\rm H}$}
were derived from the observed spectra after ensuring that none of the
37 stars are spectroscopic binaries.  In practice,
\citet{VanReeth2015} found the following ranges for these two
parameters for the sample of 37 stars: {\tt vosc$_{\rm H}$} $\in
[2.1;4.7]\,$km\,s$^{-1}$ and {\tt vsini$_{\rm H}$} $\in
[11;170]\,$km\,s$^{-1}$.  The values and ranges reveal that this
sample of bona fide $\gamma\,$Dor stars consists of slow to moderate
rotators (compared to their breakup velocity) and that their
rotational velocity is typically an order of magnitude larger than
their tangential g-mode and microturbulent velocity together, where we
recall that both quantities are integrations across the visible
stellar surface in the line-of-sight. Since these two velocity
components influence the width of spectral lines added in quadrature,
these ranges show that it is extremely challenging to unravel
pulsational from rotational broadening, even for high-resolution
spectroscopy \citep[cf.,][their Fig.\,8]{Aerts2004}. Moreover, since
snapshot spectra cannot deliver proper time-dependent line-profile
variations and only the line broadening is deduced, assuming a
symmetrical line while ignoring its true shape, the relative
uncertainties for {\tt vosc$_{\rm H}$} are considerable.

We use the 37 measured values for {\tt vbroad$_{\rm H}$}, {\tt
  $v\sin\,i_{\rm H}$}, and {\tt vosc$_{\rm H}$} to interpret the
Gaia DR3 measurements of the overall line broadening (denoted as 
{\tt vbroad$_{\rm G}$} for the bona fide $\gamma\,$Dor stars),
realising that the RVS resolving power
is in principle insufficient to unravel pulsational from rotational
line broadening. DR3 delivered {\tt vbroad$_{\rm G}$} for 27 of the bona fide
$\gamma\,$Dor stars.  We used these values to treat the following questions:
  \begin{enumerate}
  \item
    Is the quantity {\tt vbroad$_{\rm G}$} obtained by Gaia RVS different from the
    independently obtained higher-precision quantities {\tt
      vbroad$_{\rm H}$} or {\tt vsini$_{\rm H}$}?
  \item
    Can the variability in {\tt vbroad$_{\rm H}$} measured from HERMES
    for the sample of the 37 bona
    fide $\gamma\,$Dor stars be predicted by the Gaia DR3 covariates
$\log\,T_{\rm eff}$, $\log\,g$,
$\log\,(L/{\cal L}_\odot$), $\nu$, and $A_\nu$ and if so with what
    kind of quality?    
    \end{enumerate}
  Answering these two questions will help to find an astrophysical
  interpretation of Gaia's {\tt vbroad} values for the two new large
  samples of g-mode pulsators without being able to rely on
  measurements of line broadening deduced from high-resolution
  spectroscopy as we only have it available from a homogeneous data
  analysis for the 37 bona fide $\gamma\,$Dor pulsators.
  
\begin{table*}
\begin{center}
\caption{Estimates (and standard errors) for the model parameters of the
  errors-in-variables model in Eq.\,(\ref{vijf}) fitted to the bona
  fide $\gamma$\,Dor stars, for
  four combinations of $X$ and $Y$, where the HERMES quantities are
  available for all 37 stars and  {\tt vbroad$_{\rm G}$} for 27 of them.
  \label{table1}}
\begin{tabular}{lcrrrr}
\hline
&$X$& {\tt vbroad$_{\rm G}$}& {\tt vbroad$_{\rm H}$}&{\tt vsini$_{\rm
    H}$}
&{\tt vsini$_{\rm H}$} \\
&$Y$& {\tt vbroad$_{\rm H}$}& {\tt vbroad$_{\rm G}$}& {\tt
  vbroad$_{\rm H}$}
& {\tt vbroad$_{\rm G}$}\\
\hline
Effect&Par.&\multicolumn{4}{c}{Estimates (s.e.)}\\
\hline
Intercept    &$\beta_0$ &4.8(2.4)&$-$2.5(2.5)&0.53(0.97)&$-$2.3(2.4)\\
Slope        &$\beta_1$ &1.02(0.04)& 0.94(0.04)&0.99(0.02)& 0.94(0.04)\\
Res.~var.    &$\sigma^2$&0.0000(0.0002)& 0.0000(0.0000)&0.0000(0.0000)& 0.0000(0.0000)\\
\hline
Effect&Par.&\multicolumn{4}{c}{95\% confidence intervals}\\
\hline
Intercept    &$\beta_0$ &[$-$0.16;9.74]&[$-$7.63;2.55]&[$-$1.44;2.49]&[$-$7.19;2.56]\\
Slope        &$\beta_1$ &[0.94;1.11]&[0.86;1.02]&[0.95;1.03]&[0.86;1.01]\\
\hline
\end{tabular}
\end{center}
\end{table*}

To tackle the first question, we fit the statistical model in
Eqn.\,(\ref{vijf}) for four combinations of $X$ and $Y$. The parameter
estimates and statistical properties of the regression models are
presented in Table~\ref{table1}.  We find that the residual variances
$\sigma^2$ are all extremely close to zero. None of the intercepts are
significantly different from zero, and none of the slopes are
significantly different from unity. This implies that all three quantities
are essentially equal to each other, within the uncertainty limits
specified by the measurement errors. The fractions of the variance
explained by each of the four models range from 94\% to 100\%.

Given that {\tt vbroad$_{\rm G}$} is missing for 10 of the 37
$\gamma\,$Dor stars, the models involving this variable were refitted
after multiple imputation \citep{MK07} to examine the potential impact
of missingness on the results. This well-known statistical technique
was only recently introduced in astrophysics for the treatment of
missing data, such as the multivariate stellar astrophysics study
relating nine measured quantities to surface nitrogen by
\citet{Aerts2014a} and the time-series analysis of visual binaries by
\cite{Claveria2019}. The method was applied here as follows. First,
based on a so-called imputation model, 100 copies of each missing
values are drawn from the predictive distribution of what is missing,
given what is observed. Second, each so-completed dataset is then
analysed with the model that would be used had the data been
complete. Third, the 100 results combined into a single result using
appropriate combination rules. Results were qualitatively very similar
to those reported in Table~\ref{table1}, offering confidence that
missingness does not play an important role in the relationships
offered in Table~\ref{table1}. For this reason and simplicity, we
proceed with the results presented in Table~\ref{table1}.

Following up on the study by \citet{VanReeth2015}, we conclude from
the bona fide $\gamma\,$Dor g-mode pulsators that single epoch
spectra, even if of high resolution and high SNR, cannot distinguish
the overall line broadening from the line broadening caused by just
rotation when working with a fudge parameter relying on the assumption
of a time-independent symmetrical line profile. Given this spectral
line modelling limitation, we find that Gaia RVS delivers good
approximate values for spectral line broadening compared to those
deduced from snapshot high-resolution spectroscopy for early F-type
stars. Yet the uncertainties deduced from the HERMES spectra are
lower, because its better suitable spectral coverage includes more
spectral lines whose shape is determined by the temperature rather
than pressure broadening.

To address the second question, we examine the effect of the Gaia DR3
variables $\log\,(L/{\cal L}_\odot$), $\log\,T_{\rm eff}$, $\log\,g$,
$A_\nu$, and $\nu$ on the independently obtained parameter {\tt
  vbroad$_{\rm H}$}. This can be done with or without adding {\tt
  vbroad$_{\rm G}$} and with or without adding {\tt vosc$_{\rm H}$} to
the set of predictors. We proceed by backward selection, starting with
the full set of predictors and then progressively removing the one
with the highest $p$-value, until only significant effects remain
(i.e., all $p\le 0.05$).  In both versions with {\tt vbroad$_{\rm G}$}
included in the predictor set, this is the only one remaining after
model selection, and we recover the result already reported in
Table~\ref{table1}, as expected.

\tabcolsep=3.pt
\begin{table}
\begin{center}
\caption{Estimates (standard errors) for the model
  parameters of the errors-in-variables model for {\tt vbroad$_{\rm
      H}$}, fitted to the bona fide $\gamma$\,Dor stars, based on backward
  selection from a set of predictors.\label{table2}}
\begin{tabular}{lcrrr}
\hline
Effect&Par.&Est. (s.e.)&$p$-value&95\% conf. int.\\
\hline
Intercept            &$\beta_0$& 95(4) &         &[86;104]\\
$100\cdot(\log\,T_{\rm eff}\!-\!3.85)$  &$\beta_1$&$-$28(5) &$<$0.0001&[$-$37;$-$18]\\
$\log\,g-4.0$         &$\beta_2$& 60(14)&   0.0001&[32;88]\\
$\nu -1.75$        &$\beta_3$& 57(6) &$<$0.0001&[44;70]\\
Res.\ var.&           $\sigma^2$&70(152)&0.3623   &[$-$238;378]\\
\hline
\end{tabular}
\end{center}
\end{table}

When {\tt vbroad$_{\rm G}$} cannot be considered, as is the case for
the majority of Gaia targets, the following insignificant predictors
are removed by means of backward selection: first $A_\nu$, second
$\log\,(L/{\cal L}_\odot$), and third {\tt vosc$_{\rm H}$.}  Since the
latter variable is removed, whether or not it is included among the
predictors to select from does not matter. Hence, only one additional
model is obtained, the fit of which is presented in
Table~\ref{table2}. This model explains about 58\% of the variance
present in {\tt vbroad$_{\rm H}$} via the effective temperature,
surface gravity, and dominant frequency as covariates, which are all delivered
by Gaia DR3.  We note that the ranges of the covariates are
$[3.83;3.87]$ for $\log\,T_{\rm eff}$, $[3.71;4.48]$ for $\log\,g$,
and $[0.78;3.01]$\,d$^{-1}$ for $\nu$, which is why they were linearly
transformed as displayed in Table~\ref{table2} to stabilise the model
fit optimally.

In response to the second question, we find the dominant g-mode
frequency $\nu$ to be a significant predictor of the high-resolution
spectroscopic line broadening, alongside the temperature and gravity
of the star. This offers the opportunity to predict the line
broadening for all the Gaia $\gamma\,$Dor stars without a Gaia
measurement of the line broadening if these three covariates are
available, as is the case for the majority of the Gaia DR3
$\gamma\,$Dor stars. Of course it should be born in mind that some
predictors exhibit mild to strong correlation, given their
astrophysical meaning. In the particular application of the bona fide
$\gamma\,$Dor stars, the strongest correlation among the predictors is
the one between $\log\,(L/{\cal L}_\odot)$ and $\log\,g$, namely
-0.70. The correlation between $\log\,T_{\rm eff}$ and $A_\nu$ is 0.40,
while $\nu$ correlates equally with $\log\,T_{\rm eff}$ and with
$\log\,g$ with a moderate value of 0.32. All other correlations are
much smaller. Hence, the regression
coefficients in a model with multiple predictors should be interpreted
as the effect of change by one unit in a predictor, while all others
remain constant, in this case for the three surviving predictors
$\log\,T_{\rm eff}$, $\log\,g$, and $\nu$.  A graphical perspective on
the predictions for {\tt vbroad$_{\rm H}$} from {\tt vbroad$_{\rm
    G}$}, {\tt vsini$_{\rm H}$}, and the three covariates is shown in
Fig.\,\ref{Geert}, using Eqns.\,(\ref{zeven})--(\ref{acht}).

\subsection{Results for the Gaia DR3 $\gamma\,$Dor pulsators}

\begin{figure*}[h!]
\begin{center} 
\rotatebox{270}{\resizebox{6.1cm}{!}{\includegraphics{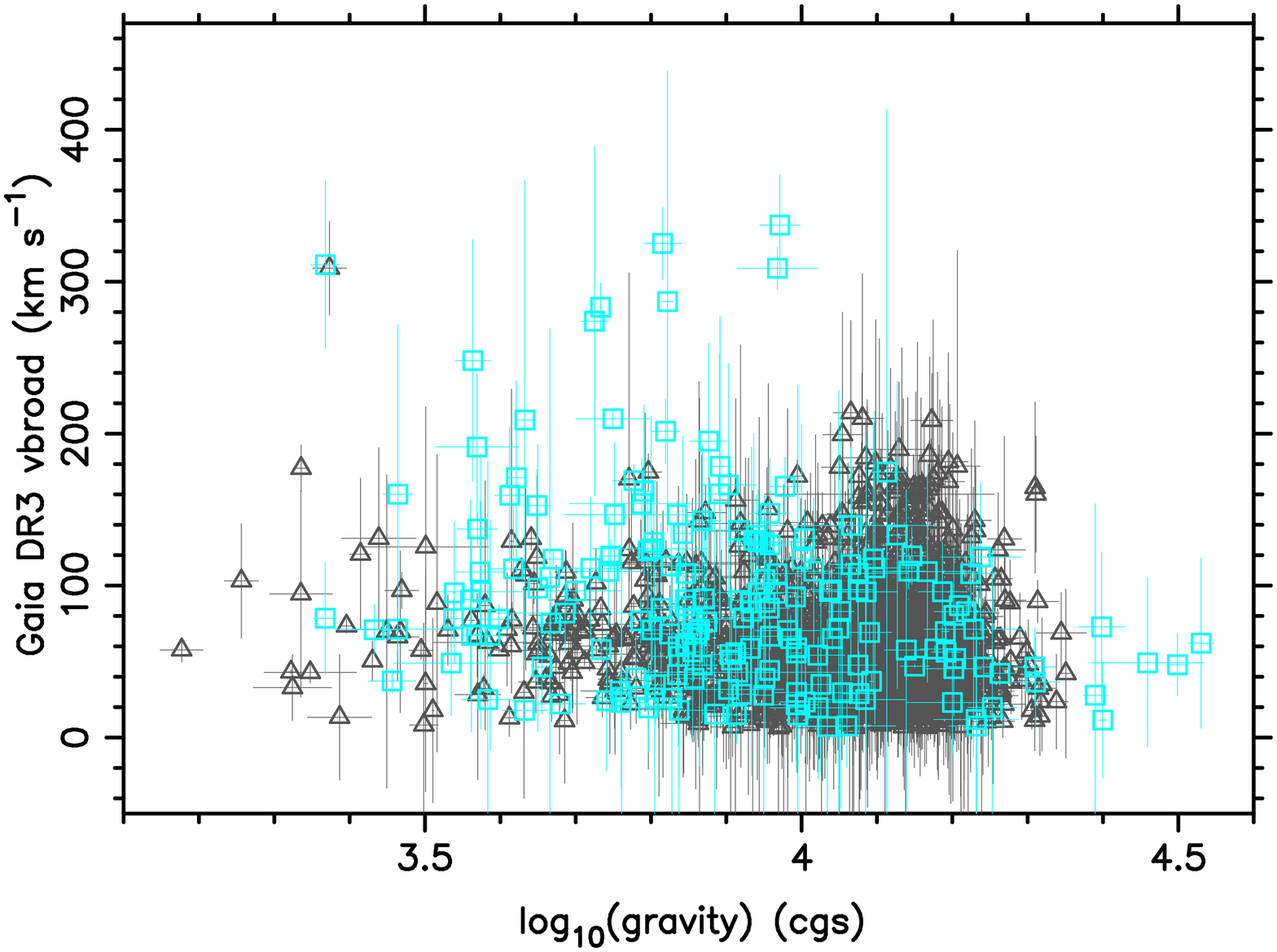}}}\hspace{0.5cm}
\rotatebox{270}{\resizebox{6.1cm}{!}{\includegraphics{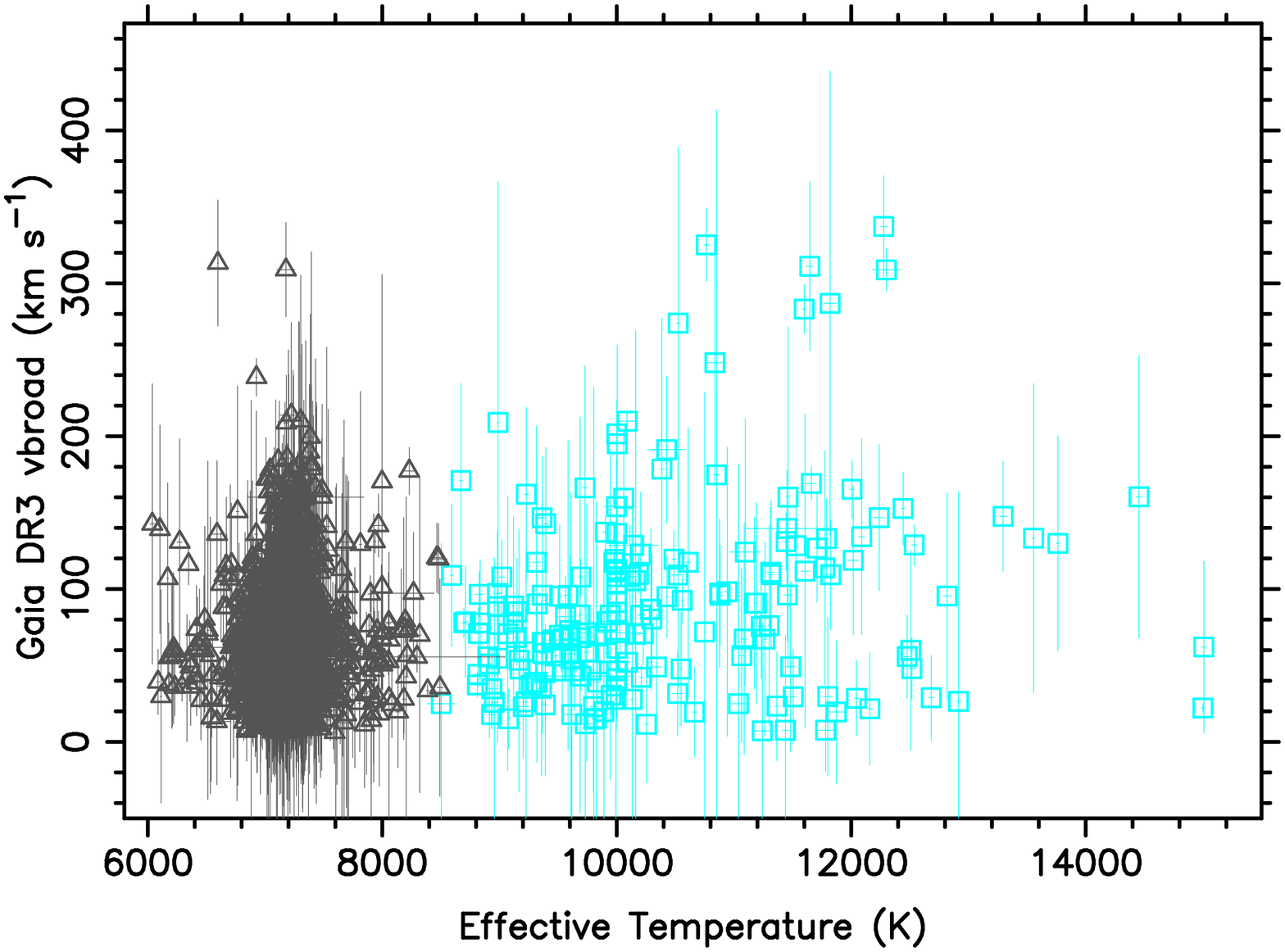}}}\vspace{0.5cm}
\rotatebox{270}{\resizebox{6.1cm}{!}{\includegraphics{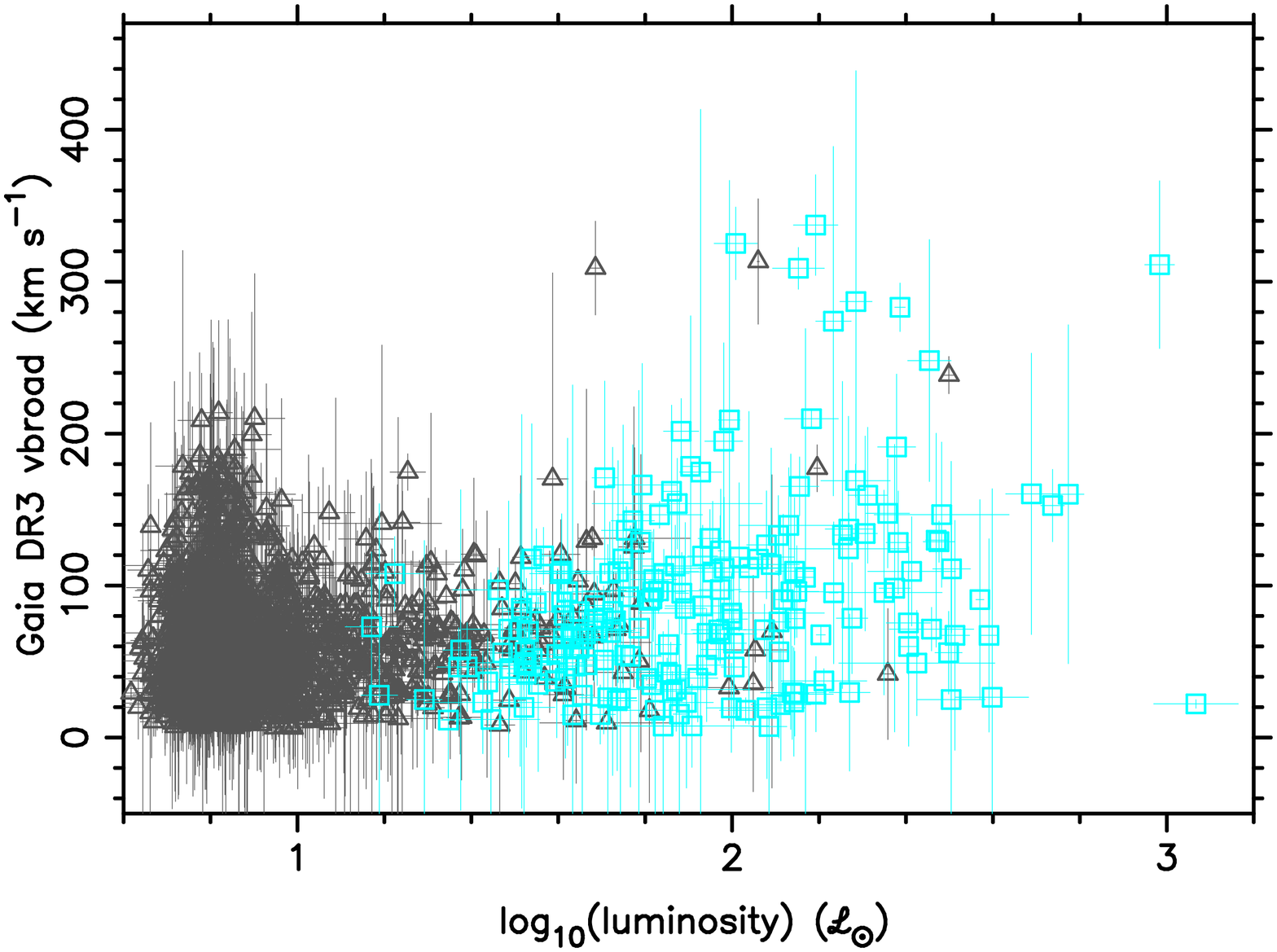}}}\hspace{0.5cm}
\rotatebox{270}{\resizebox{6.1cm}{!}{\includegraphics{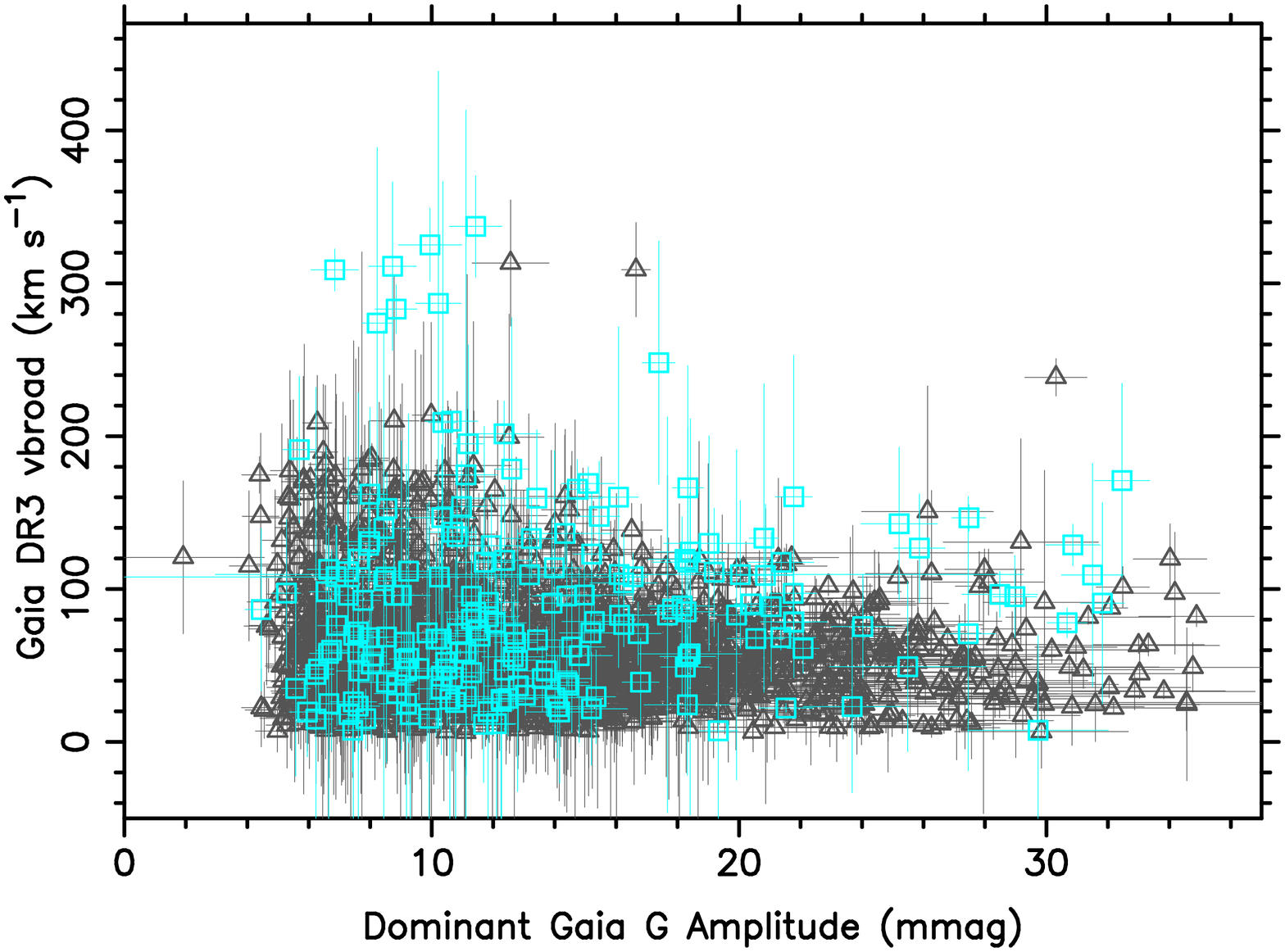}}}\vspace{0.5cm}
\rotatebox{270}{\resizebox{6.1cm}{!}{\includegraphics{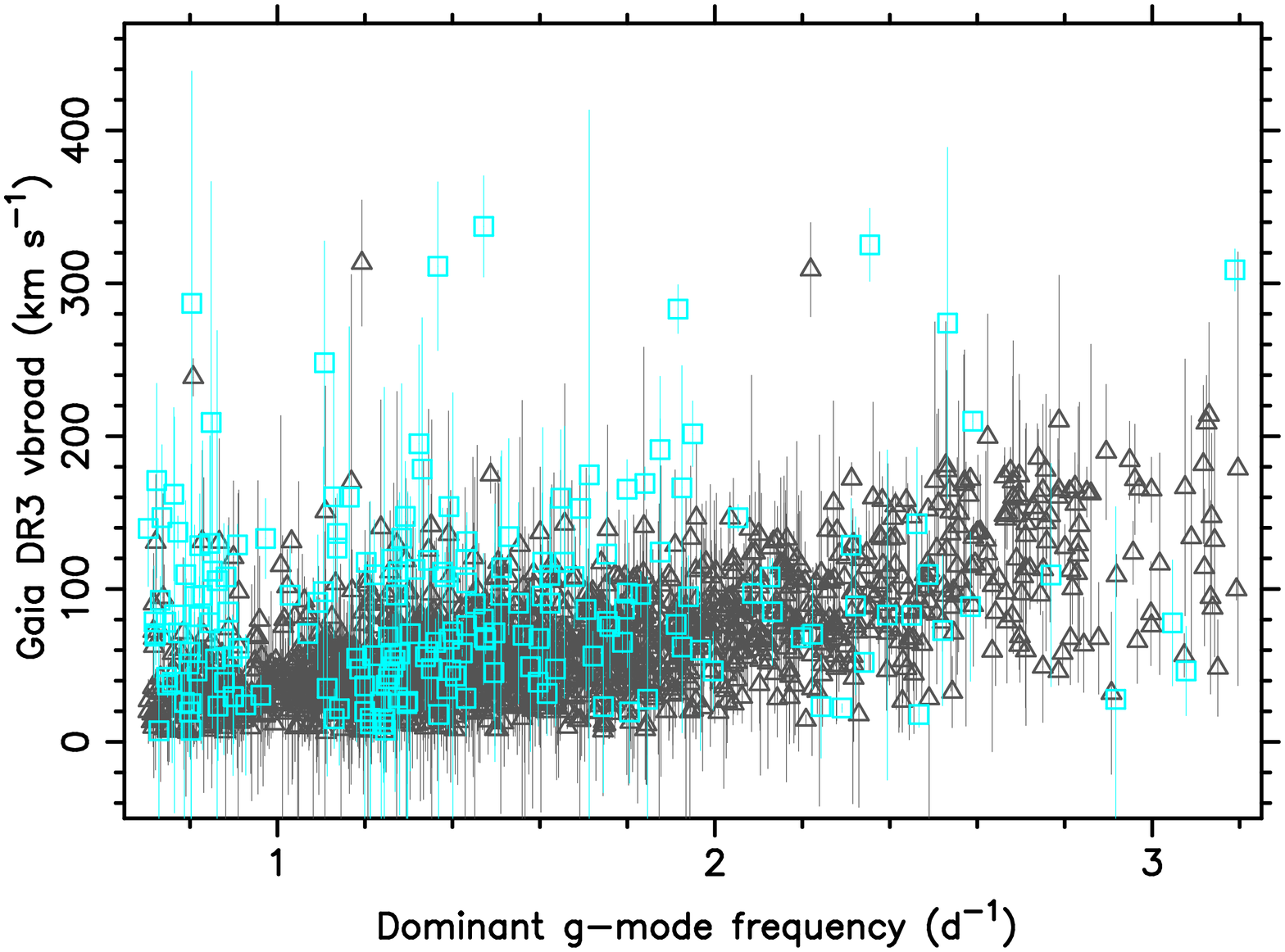}}}\hspace{0.5cm}
\rotatebox{270}{\resizebox{6.1cm}{!}{\includegraphics{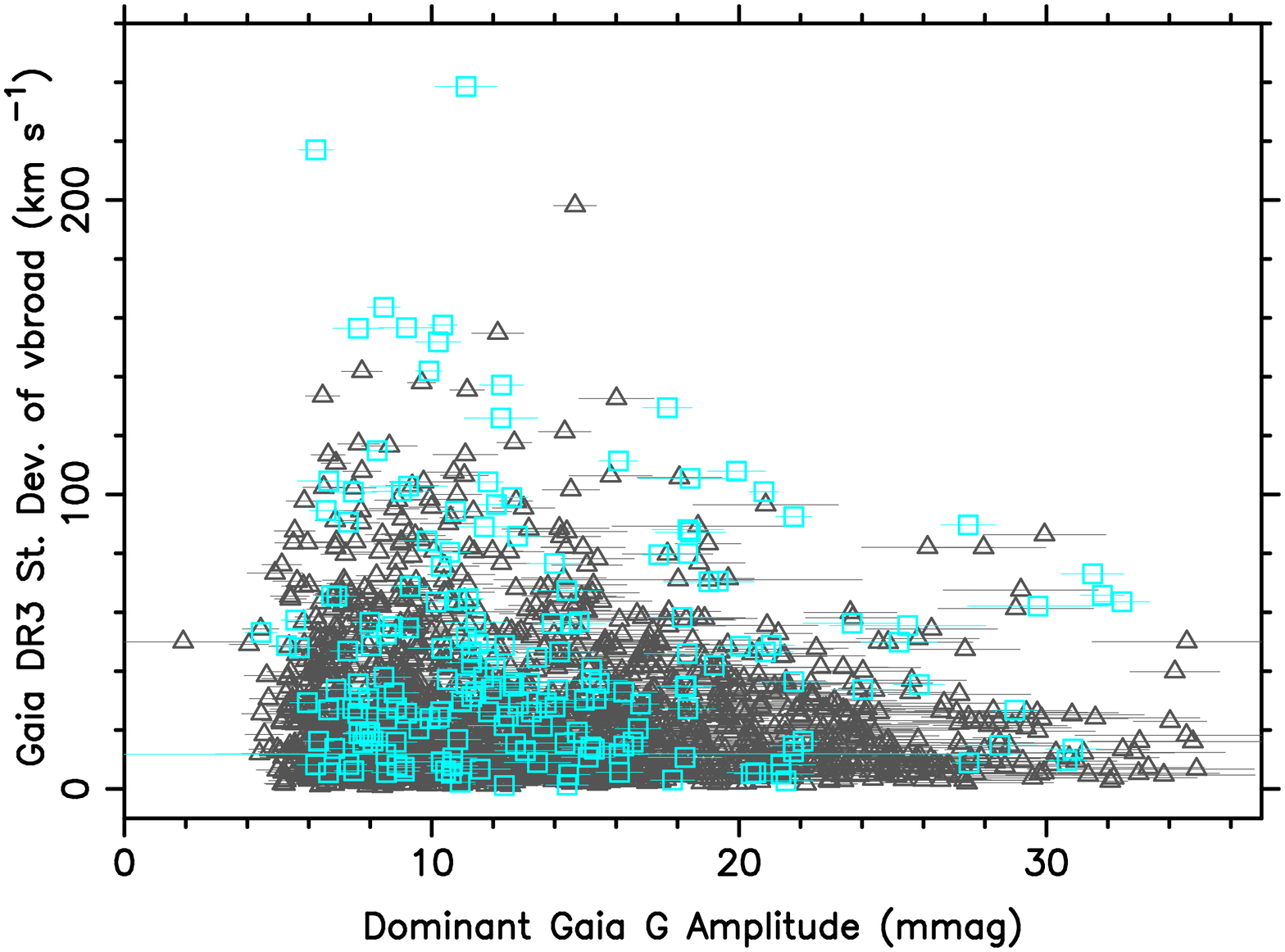}}}
\end{center}
\caption{\label{covariates} Gaia DR3 measurements of {\tt vbroad}
  versus each of the five covariates as indicated for the {  1775}
  $\gamma\,$Dor (grey triangles) and {  190} SPB (cyan squares) stars
  having these quantities available. The lower right panel shows the
  standard deviation of {\tt vbroad} as a function of the dominant
  g-mode amplitude. When invisible, the errors are smaller than the
  symbol sizes.}
\end{figure*}

Armed with the knowledge that {\tt vbroad$_{\rm G}$} and {\tt
  vbroad$_{\rm H}$} are equal for the 37 bona fide $\gamma\,$Dor stars
to within their measurement uncertainties from high-resolution and
Gaia RVS spectra, and with the predictive model for these quantities
given in Table~\ref{table2}, we now look at the sample of
{  11\,636} Gaia
DR3 $\gamma\,$Dor stars. For all of those, full information is
available on $\log\,T_{\rm eff}$, $\log\,g$, $\log\,(L/{\cal
  L}_\odot)$, $\nu$, and $A_\nu$, deduced in one homogeneous way from
Gaia DR3 following Paper\,I.  For {  100} of these stars, both {\tt
  vbroad$_{\rm G}$} (as of now again simplified to {\tt vbroad}) and
{\tt vsini$_{-}$esphs} are recorded. These are the so-called
completers of the Gaia DR3 $\gamma\,$Dor stars.  For
{  1775}
$\gamma\,$Dor stars there is a measurement on {\tt vbroad} but not on
{\tt vsini$_{-}$esphs} and for {  384} {\tt vsini$_{-}$esphs} information
is available but {\tt vbroad} is missing. For the remaining
{  9577} stars both of these line broadening quantities are missing.

\tabcolsep=3.5pt
\begin{table}
\begin{center}
\caption{Estimates
  (standard errors) for the model parameters of the
  errors-in-variables model, relating {\tt vbroad}  to {\tt
    vsini$_{-}$esphs}, on the {  100} completers within the Gaia DR3 set of
  $\gamma$\,Dor stars.
  \label{vbroadgvsinig}}
\begin{tabular}{lcrrr}
\hline
Effect&Par.&\multicolumn{1}{c}{Estimate
  (s.e.)}&$p$-value&\multicolumn{1}{c}{95\% conf. int.}\\
\hline
Intercept    &$\beta_0$ &$-$5(3)&&[$-$12;2]\\
Slope        &$\beta_1$ &0.90(0.05)&$<$0.0001&[0.81;0.99]\\
Res.~var.    &$\sigma^2$&0(4)&0.98&[$-$7;7]\\
\hline
\end{tabular}
\end{center}
\end{table}

When considering the {  100} completers only, we again conclude that {\tt
  vbroad} and {\tt vsini$_{-}$esphs} are identical within the bounds
specified by the measurement errors, given that the slope parameter is
roughly equal to unity and the residual variance is not different from
zero (see Table\,\ref{vbroadgvsinig}).  Figure\,\ref{vbroad-vsini}
already showed that the two variables {\tt vbroad} and {\tt
  vsini$_{-}$esphs} are similar for the $\gamma\,$Dor stars with
both estimates, while graphically revealing the different meaning of
their uncertainty regions. The grey dashed line in that figure
represents the regression model in Table~\ref{vbroadgvsinig}.

For the full Gaia DR3 $\gamma\,$Dor data set, the relationship between
{\tt vbroad} and {\tt vsini$_{-}$esphs} can also be assessed using all
stars after performing multiple imputation. For this application we
drew ten imputations using information on {\tt vbroad}, {\tt
  vsini$_{-}$esphs}, their standard errors, and the covariates
$\log\,T_{\rm eff}$, $\log\,g$, $\log\,(L/{\cal L}_\odot)$, $\nu$, and
$A_\nu$. Given that {  less than 1\%} of stars have complete
information, the relationship found from multiple imputation is
relatively different from the one found for the {  100} completers. This
is not surprising given the relatively large uncertainties of the two
broadening parameters and the fact that they are rather weakly
correlated with other information in the dataset.  Moreover, the very
large fraction of incomplete information destabilizes the inference
from multiple imputation. These issues taken together suggest that the
results of multiple imputation are too unstable to provide trustworthy
results. For these reasons, further analysis will be based on
completers only for each of the regression applications discussed
below.

We now turn to the relationship between {\tt vbroad} and the predictor
variables, applying backward selection to the errors-in-variables
model for the {  1775} $\gamma\,$Dor stars having this quantity and the
covariates $\log\,T_{\rm eff}$, $\log\,g$, $\log\,(L/{\cal L}_\odot$),
$\nu$, and $A_\nu$ (whose regression coefficients we denote as
$\beta_1, \beta_2, \beta_3, \beta_4, \beta_5$, respectively).  It
turns out that all these covariates are significant except the
amplitude of the dominant frequency, which has a $p-$value of {  0.0662}
and is thus borderline significant.  This is why we present the
regression models with and without this covariate in
Table~\ref{vbroadGcov}. Both these models explain {  42\%} of the variance
in the measurements of {\tt vbroad}. It can be seen in
Table~\ref{vbroadGcov} that keeping $A_\nu$ in the model does not
alter the regression coefficients of the other four covariates. We
show the measurements of {\tt vbroad} as a function of each of the
covariates in Fig.\,\ref{covariates}.

\begin{table}
\begin{center}
\caption{Estimates (standard errors) for the parameters of the
  errors-in-variables model for {\tt vbroad} for the {  1775} Gaia DR3
  $\gamma\,$ Dor stars with measured values for this quantity, based
  on backward selection from the set of listed predictors. Both models
  explain 42\% of the variance in {\tt vbroad}. \label{vbroadGcov}}
\begin{tabular}{lcrrr}
\hline
Effect&Par.&Estimate (s.e)&$p$-value&95\% conf. int.\\
\hline
\multicolumn{5}{c}{With $A_\nu$}\\
\hline
Intercept            &$\beta_0$&$-$478(262)&&[$-$993;$-$36]\\
$\log\,T_{\rm eff}$   &$\beta_1$& 238(87)&0.0065&[67;409]\\
$\log\,g$           &$\beta_2$& $-$100(19)&$<$0.0001&[$-$138;$-$63]\\
$\log\,(L/{\cal L}_\odot$) & $\beta_3$& $-$45(15)&0.0022&[$-$74;$-$16]\\
$\nu$              &$\beta_4$&44(1)&$<$0.0001&[41;46]\\
$A_\nu$             &$\beta_5$&$-$192(104)&0.0662&[$-$397;13]\\
Res.\ var.           &$\sigma^2$&397(20)&$<$0.0001&[358;437]\\
\hline
\multicolumn{5}{c}{Without $A_\nu$}\\
\hline
Intercept            &$\beta_0$& $-$478(260)&&[$-$988;$-$32]\\
$\log\,T_{\rm eff}$   &$\beta_1$&237(87)&0.0062&[68;407]\\
$\log\,g$                 &$\beta_2$&$-$100(19)&$<$0.0001&[$-$138;$-$63]\\
$\log\,(L/{\cal L}_\odot$) & $\beta_3$& $-$45(15)&0.0023&[$-$73;$-$16]\\
$\nu$                 &$\beta_4$&44(1)&$<$0.0001&[41;47]\\
Res.\ var.           &$\sigma^2$&395(20)&$<$0.0001&[356;435]\\
\hline
\end{tabular}
\end{center}
\end{table}

As discussed in Sect.\,5.1, {\tt vsini$_{-}$esphs} is based on the
averaged BP/RP (and averaged RVS spectrum when available) and
therefore has smaller uncertainty than {\tt vbroad} whose uncertainty
interval represents the time-dependent line broadening covered by at
least six snapshot spectra. By construction, {\tt vsini$_{-}$esphs} is
expected to be a better representative of the time-independent surface
rotation velocity of the star than {\tt vbroad}. Indeed, the latter
quantity approximates the overall time-dependent spectral line broadening
due to various phenomena acting together because it was computed as
the median value from individual transits taken at different epochs
and treated as such by the MTA.

To test if {\tt vsini$_{-}$esphs} and {\tt vbroad} indeed
capture different astrophysical information, we repeat the same
backward model selection for {\tt vsini$_{-}$esphs},
considering the same covariates for the {  384} $\gamma\,$Dor stars having
a measurement of {\tt vsini$_{-}$esphs}.  This leads to the
{  
successive removal of $A_\nu$, $\log(L/{\cal L}_\odot)$, and
$\log\,T_{\rm eff}$
}
as being insignificant. The coefficients of the resulting regression
model are listed in Table~\ref{vsiniGcov}, while the plots of 
 the measurements of {\tt vsini$_{-}$esphs} as a function of each of the
 covariates are included in Appendix\,A (Fig.\,\ref{covariatesvsini},
 to be compared with Fig.\,\ref{covariates}).
We find that {\tt
  vsini$_{-}$esphs} does not depend on the effective temperature and
the luminosity, while the surface gravity and dominant frequency remain
significant covariates. These two covariates offer the same dependence
for {\tt vsini$_{-}$esphs} as for {\tt vbroad}, that is lower $\log\,g$
(a more evolved star) and higher $\nu$ give larger line broadening.
As far as $\nu$ is concerned, this is well understood in terms of an
asteroseismic interpretation and in agreement with the findings based
on the HERMES spectroscopy by \citet{VanReeth2015} for the bona fide
$\gamma\,$Dor stars.  Indeed, a higher dominant g-mode frequency in
the inertial frame of an observer corresponds to a faster rotating
star \citep[][for galleries of {\it Kepler\/} light curves and
  frequency spectra as a function of rotation
  frequency]{VanReeth2015,VanReeth2016,Papics2017}.  Hence, higher
asteroseismic $\nu$ is a signature of faster stellar rotation and thus
of larger line broadening, irrespective of whether one considers {\tt
  vsini$_{-}$esphs} or {\tt vbroad}.

While the resulting regression model for {\tt vsini$_{-}$esphs} of
{  384} 
class members explains only {  5\%} of the variance in that quantity, it
is {  42\%} for {\tt vbroad} of the {  1775} stars having this quantity. Thus
the time-independent projected rotational velocity represented by {\tt
  vsini$_{-}$esphs} of the $\gamma\,$Dor stars is independent of their
effective temperature and luminosity, while inversely proportional to
(but only weakly dependent on) their gravity. On the other hand, the
time-dependent quantity {\tt vbroad} does connect to the effective
temperature of the $\gamma\,$Dor stars, such that the hotter the star,
the larger {\tt vbroad}. Our astrophysical interpretation of these
findings connects well to the excitation mechanisms and to the level of
line broadening found for $\gamma\,$Dor stars in the
literature. Indeed, since the {\it Kepler\/} data allowed for detailed
asteroseismic modelling, we know that the dominant modes of the bona
fide g-mode pulsators are dipole prograde modes and that these stars
occupy a narrow range in mass, namely $[1.3;1.9]\,$M$_\odot$, while
they cover the entire main sequence \citep{Mombarg2019,Mombarg2021}.
This pulsation class thus has stars with quite a broad range of
$\log\,g$ and radii (cf.\,Fig.\,\ref{hist-radius}). The variability in
$\log\,T_{\rm eff}$ and $\log\,g$ revealed among the class members is
thus mainly a signature of evolutionary status. 

The regression models for {\tt vbroad} and {\tt vsini$_{-}$esphs}
reveal more evolved stars to have larger spectral line broadening,
while maximal time-dependent line broadening occurs for $T_{\rm eff}$
between 6500 and 7500\,K (cf.\,the grey triangles in the upper left
panel of Fig.\,\ref{covariates}). This is precisely the temperature
range where \citet{Grassitelli2015b} found a maximal effect of
turbulent pressure in the stellar envelope of evolved A- and F-type
dwarfs, offering an additional mechanism to excite high-order
eigenmodes in such objects, aside from the classical
$\kappa$\,mechanism being active in the hotter $\gamma\,$Dor stars and
flux blocking at the bottom of the convective envelope causing such
g~modes in the cool class members
\citep{Guzik2000,Dupret2005,Xiong2016}.  In addition,
\citet{Tkachenko2020} already showed that ignoring the turbulent
pressure in stellar atmosphere models affects estimation of
microturbulent broadening and results in an overestimation of the
effective temperature at a few \% level. Moreover, the authors found
this effect to get worse as the star evolves, that is for decreasing
$\log\,g$.  We thus conclude to have found observational evidence from
Gaia DR3 {\tt vbroad} measurements that time-dependent macroturbulent
spectral line broadening in these stars is connected with their
excited g~modes and/or surface gravity, in addition to surface
rotation. The amplitude limitation from Gaia DR3 and the comparative
distributions of the dominant amplitudes and frequencies between the
Gaia DR3 and bona fide $\gamma\,$Dor pulsators (cf.\,left panels of
Figs\,\ref{hist-freq} and \ref{hist-ampl}) suggest that the detected
dominant frequencies are due to large-scale (i.e., low-degree)
gravito-inertial modes. The interplay of the dominant g~mode with the
rotation of the star, along with variability in $\log\,T_{\rm eff}$
and $\log\,g$ due to poor treatment of turbulent pressure in the
{  line-forming region}, explain the overall spectral line broadening
estimates from Gaia DR3.

We point out that the regression
model for {\tt vbroad} in Table\,\ref{vbroadGcov} explains
{  42\%} of the
measured variance in the spectral line broadening, while it was 58\%
for the bona fide pulsators. Both these results are readily understood
given that we are dealing with multiperiodic g-mode pulsators.
Indeed, $\gamma\,$Dor pulsators have tens of high-order low-degree
g~modes active simultaneously, whichever of the three excitation
mechanisms is dominant \citep{VanReeth2015}. The line broadening
captures the collective effect of all these modes together
\citep{Aerts2009}.  Yet, the frequencies of the excited g~modes in
addition to the dominant one were not included in the regression
model, because the Gaia light curves currently do not provide
sufficient data to unravel the multiperiodic oscillations active in
these stars. While the frequencies and amplitudes of the second
strongest variability signal were determined in Paper\,I, it was found
that quite a number of those frequencies cannot be distinguished from
frequencies above 3\,d$^{-1}$ that may result from instrumental
effects. That is why we did not use these secondary frequencies from
DR3.  It is to be anticipated that improved regression models
explaining a higher fraction of the variance in the spectral line
broadening will become possible from DR4 and particularly DR5, because
the longer time base and doubling of the number of epochs in the Gaia
photometry will allow to unravel several additional g-mode frequencies,
particularly when combined with additional light curves dedicated to
asteroseismology as illustrated from combined Hipparcos and TESS or
ground-based data \citep[cf.\,][]{Waelkens1998,DeCat2007,Cuypers2009}.
Yet, Gaia's sampling is too sparse to deliver all the modes active in these
multiperiodic g-mode pulsators, while they do contribute to the overall
broadening of the spectral lines \citep[cf.\,][for the theoretical
  expression of the spectral line width due to multiperiodic non-radial
  oscillations]{Aerts1994}.  The fraction of the variance explained by
regression models relying on the fundamental parameters and the
significant frequencies in Gaia light curves will therefore always be
limited, even for the bona fide class members. In this sense, the { 
  42\%}
reached for the model in Table\,\ref{vbroadGcov} is high.

\begin{table}
\begin{center}
\caption{Estimates
  (standard errors) for the parameters of the
  errors-in-variables model for {\tt vsini$_{-}$esphs} for the {  384}  Gaia
  DR3 $\gamma\,$ stars with measured values for this quantity, based on
  backward selection from a set of predictors.\label{vsiniGcov}}
\begin{tabular}{lcrrr}
\hline
Effect&Par.&Estimate (s.e.)&$p$-value&95\% conf. int.\\
\hline
Intercept            &$\beta_0$&198(56)&&[88;308]\\
$\log\,g$                 &$\beta_2$&$-$37(15)&0.0161&[$-$66;$-$7]\\
$\nu$            &$\beta_4$&17(6)&0.0047&[5;28]\\
Res.\ var.           &$\sigma^2$&599(152)&0.0002&[296;901]\\
\hline
\end{tabular}
\end{center}
\end{table}


\subsection{Results for the Gaia DR3 SPB pulsators}

We now repeat the same analyses for the {  3426} new Gaia DR3 SPB stars.  Among these,
both {\tt vbroad}  and {\tt vsini$_{-}$esphs} are recorded
for {  156} stars. For {  34} of them there is a measurement on {\tt
  vbroad} but not on {\tt vsini$_{-}$esphs}, while for {  948} 
{\tt vsini$_{-}$esphs} is available but {\tt vbroad}  is
missing. For the remaining {  2288} SPB stars both of these are missing.
In line with the arguments provided in Section\,5.4,
attention is restricted to an analysis of the completers to
test relationships for {\tt vsini$_{-}$esphs} and {\tt vbroad}. 

\begin{table}
\begin{center}
\caption{Estimates (standard errors) for the model parameters of the
  errors-in-variables model, relating {\tt vbroad}  to {\tt
    vsini$_{-}$esphs}, on {  156} completers within the SPB sample.
  \label{vbroadgvsinig3}}
\begin{tabular}{lcrrr}
  \hline
Effect&Par.&\multicolumn{1}{c}{Estimate
  (s.e.)}&$p$-value&\multicolumn{1}{c}{95\% conf. int.}\\
\hline
Intercept    &$\beta_0$ &-11(5)&&[$-$20;$-$2]\\
Slope        &$\beta_1$ &1.06(0.05)&$<$0.0001&[0.97;1.15]\\
Res.~var.    &$\sigma^2$&25(30)&0.41&[$-$35;84]\\
\hline
\end{tabular}
\end{center}
\end{table}

The results of the test whether or not the two measures for the spectral line
broadening are equal are given in Table\,\ref{vbroadgvsinig3} and shown
graphically in Fig.\,\ref{vbroad-vsini} (cyan symbols).  Just like for
the Gaia DR3 $\gamma\,$Dor sample, we again have a relationship that
is consistent with the hypothesis that {\tt vbroad} and {\tt
  vsini$_{-}$esphs} are identical, keeping in mind the errors for {\tt
  vsini$_{-}$esphs} and the standard deviation for {\tt vbroad}.

Next, the relationship between {\tt vbroad} and the candidate
predictors is examined relying on the {  190} SPB stars having these data,
once again applying backward selection to the errors-in-variables
model. {  All covariates are again significant.} The
regression model in Table\,\ref{vbroadGcov3} explains {  21\%} of the
variance in {\tt vbroad}. Backward selection applied to {\tt
  vsini$_{-}$esphs} for the {  1104} SPB stars with this quantity reveals
{  $T_{\rm eff}$} and $\nu$ to be significant predictors, with a
regression model explaining 8\% of the variance.
For the regression coefficient of $\nu$, we assign the same
astrophysical interpretation as before that higher $\nu$ corresponds to faster
surface rotation following the SPB studies by
\citet{Papics2017,Pedersen2021,Pedersen2022b}.

As for the role of the effective temperature in {\tt vbroad}
{  and {\tt vsini$_{-}$esphs}},
\citet{Papics2017} already found evidence that hotter SPB stars are
more massive and tend to rotate faster.  The $T_{\rm eff}$ dependence
revealed 
is thus in first instance a dependence on stellar mass rather
than stellar evolution as we found for the $\gamma\,$Dor stars.  This
result is as expected, given that SPB stars cover a factor 3 in mass,
from 3\,M$_\odot$ to 9\,M$_\odot$.  \citet{Pedersen2021} placed the
observed properties of the 26 bona fide {\it Kepler\/} SPB stars
included here and those studied from the ground by
\citet{DeCatAerts2002} into the context of stellar evolution theory.
This showed large diversity of $\nu$ and $A_\nu$ in terms of the
spectroscopic $\log\,T_{\rm eff}$ and $\log\,g$, as well as
$\log(L/{\cal L}_\odot)$ from Gaia DR2. This diversity was interpreted
as due to the range in mass and rotation rate, the latter covering
from almost zero to almost critical rotation for the {\it Kepler\/}
sample \citep{Aertsetal2021,Pedersen2022b}.  Despite the limited
predictive power of regression model in Table\,\ref{vbroadGcov3}, it
reveals that larger line broadening occurs for hotter {  and/or more
  evolved} SPB stars with
higher dominant g-mode frequencies (cf.\,Fig.\,\ref{covariates}). This
highlights that hotter {  younger} stars have faster rotation, shifting their
g~modes further into the gravito-inertial regime of the frequency
spectrum than those of slower rotators \citep[see][for a discussion of
  the various frequency regimes of waves connected to the dominant
  restoring forces]{Aerts2019}.

\begin{table}
\begin{center}
\caption{Estimates (standard errors) for the model parameters of the
  errors-in-variables model for {\tt vbroad}  of the {  190} SPB
  stars having a measurement for this quantity, based on
  backward selection from the set of predictors. The model explains
  {  21\%} of the variability in the data.\label{vbroadGcov3}}
\begin{tabular}{lcrrr}
\hline
Effect&Par.&Estimate (s.e.)&$p$-value&95\% conf. int.\\
\hline
Intercept            &$\beta_0$&$-$2764(643)&&[$-$4033;$-$1495]\\
$\log\,T_{\rm eff}$       &$\beta_1$&963(219)&$<$0.0001&[532;1395]\\
$\log\,g$                 &$\beta_2$&$-$213(51)&$<$0.0001&[$-$312;$-$113]\\
$\log(L/{\cal L}_\odot)$   &$\beta_3$&$-$109(42)&0.0105&[$-$192;$-$26]\\
$\nu$                 &$\beta_4$&34(8)&0.0001&[19;50]\\
$A_\nu$             &$\beta_5$&$-$1492(715)&0.0384&[$-$2903;$-$81]\\
Res.\ var.           &$\sigma^2$&1909(300)&$<$0.0001&[1317;2501]\\
\hline
\end{tabular}
\end{center}
\end{table}

\begin{table}
\begin{center}
\caption{Estimates (standard errors) for the model parameters of the
  errors-in-variables model for {\tt vsini$_{-}$esphs} based on the
  {  1104} SPB stars in the sample having a measurement for this quantity,
  based on backward selection from a set of
  predictors. The model explains 8\% of the variability in the data.\label{vsiniGcov3}}
\begin{tabular}{lcrrr}
\hline
Effect&Par.&Estimate (s.e.)&$p$-value&95\% conf. int.\\
\hline
Intercept            &$\beta_0$&$-$573(319)&&[$-$1204;58]\\
$\log\,T_{\rm eff}$       &$\beta_1$&157(80)&0.0496&[0;315]\\
$\nu$                 &$\beta_4$&20(6)&0.0016&[8;32]\\
Res.\ var.           &$\sigma^2$&745(154)&$<$0.0001&[442;1049]\\
\hline
\end{tabular}
\end{center}
\end{table}

The luminosity, $\log(L/{\cal L}_\odot)$, now also has a significant
contribution as predictor for {\tt vbroad} with a $p-$value of
{  0.0105}. The model reveals that less luminous SPB stars have higher
line broadening but its regression coefficient is not very
accurate. Moreover, the sample of SPB stars with a measurement of {\tt
  vbroad} is an order of magnitude smaller than for the $\gamma\,$Dor
stars and is skewed towards low-mass class members, limiting this
interpretation to only a small part of the SPB instability region.
This is graphically illustrated in Fig.\,\ref{covariates}, where
trends reveal that more evolved and more luminous SPB stars have
higher {\tt vbroad} but are not well enough represented in membership
to have an equally important effect on the regression model than the
cool class members.  Moreover, the luminosity panel in
Fig.\,\ref{covariates} reveals more of a quadratic than linear trend
for the SPB stars.  As already highlighted above and unlike for the
$\gamma\,$Dor stars, the luminosity of SPB pulsators is mostly
determined by their mass rather than by their evolutionary stage as
for the $\gamma\,$Dor stars.  This, along with the relatively large
scatter for $\log\,g$ and for the effective temperature, as well as
the lower fraction of the variance explained by the linear regression
model for {\tt vbroad}, makes the distillation of a simple
astrophysical interpretation for {\tt vbroad} more difficult for SPB
stars.  This conclusion is by itself fully in line with the diversity
in pulsational behaviour occurring in the {\it Kepler\/} sample of
bona fide SPB pulsators \citep{Pedersen2021,Pedersen2022a}.

Finally, we stress that time-dependent macroturbulent spectral line
broadening due to the g~modes of SPB pulsators has already been found
in several of the brightest class members \citep{Aerts2014b}, with
values in agreement with those we find here for the new faint Gaia DR3
class members. Moreover, the density of modes excited by the
$\kappa\,$mechanism peaks in the lower part of the instability strip,
near 13\,000\,K \citep{Papics2017}.  These pulsators have a mass
regime where the interpretation of turbulent pressure exciting extra
high-order g~modes in addition to the classical $\kappa\,$mechanism
does not hold \citep{Grassitelli2015a}. Macroturbulence in
these pulsators has already been established as a time-independent
downgraded quantity representing their dominant tangential pulsational
velocity by \citet{Aerts2009} and \citet{Aerts2014a}.

\section{Discussion and conclusions}

Thanks to the homogeneous treatment of its multitude of observations
and its large scale survey capacity, the Gaia mission has its role to
play for gravito-inertial asteroseismology. First of all, its
photometric light curves allow to discover thousands of new g-mode
pulsators belonging to the classes of the F-type $\gamma\,$Dor stars
or SPB-type stars.  Secondly, its broadening parameter {\tt vbroad}
contains astrophysical information on stellar oscillations having
mmag-level observed amplitudes. We found those results after
reassigning a fraction
{  of 22\% of the $\gamma\,$Dor candidates as
  SPB pulsators according to their effective temperature
being above 8500\,K, a property not taken
  into account in the variability classifications used in Paper\,I.}

We find the two samples of new Gaia DR3 g-mode pulsators to have
similar fundamental parameters than those of bona fide class members,
although the Gaia SPB pulsators only cover the cooler and less massive
class members. We studied the astrophysical properties of the new
$\gamma\,$Dor and SPB pulsators from regression models built upon the
principle of errors-in-variables, with their fundamental parameters
and dominant oscillation properties as predictors of the overall
spectral line broadening. The Gaia DR3 quantity {\tt vsini$_{-}$esphs}
offers a good estimate of the overall time-independent spectral line
broadening, reflecting that the surface rotation of the stars in our
samples is the dominant line broadening mechanism. All regression
models revealed the dominant g-mode frequency to be a significant
predictor of the Gaia DR3 {\tt vbroad} parameter and its standard
deviation, which together represent the overall time-dependent
spectral line broadening.

{  We explicitly checked via re-analyses of all regression models
  that none of the astrophysical interpretations change if we use the
  effective temperature of 9500\,K as treshold for the
  reclassifications among the $\gamma\,$Dor and SPB candidates. Such a
  treshold temperature follows from instability computations by
  \citet{Szewczuk2017} for the cool border of galactic rotating SPB
  stars instead of the adopted 8500\,K based on the hot border for the
  $\gamma\,$Dor instability strip by \citet{Xiong2016}. We find from
  the upper left panel of Fig.\,\ref{covariates} that Gaia DR3 shows
  9500\,K to be a less natural and more abrupt treshold temperature
  between the two
  classes than the adopted 8500\,K.
  Nevertheless, using 9500\,K still gives compliance of
  class populations with the IMF and almost all the coefficients obtained for
  the regression models remain within the uncertainty ranges listed
  in the tables we provided here using 8500\,K as treshold.}

Despite the limiting resolution of the RVS spectroscopy, the line
broadening of rotating g-mode pulsators offered by Gaia is in full
agreement with results of well-known class members observed with
high-precision space photometry and high-resolution spectroscopy.  In
particular, we find the dominant g-mode frequency to be a significant
predictor of the overall line broadening. This supports earlier
findings that macroturbulence is merely a simplified time-independent
approximation for the true velocity fields at the stellar surface that
cause line-profile variability {  for g-mode pulsators}
\citep{Aerts2014b}.
We conclude that
the combined effect of surface rotation and tangential velocities
resulting from multiperiodic g~modes can be estimated from {\tt
  vbroad} for the case of main-sequence stars of intermediate mass.
The regression models for {\tt vbroad} are fully in line with
{  various} excitation predictions for g~modes
in $\gamma\,$Dor and SPB
pulsators.

Given that the regression models
{  based on the fundamental parameters and on the dominant pulsation mode}
presented in Sect.\,5 explain part of
the variability of {\tt vbroad}, it is sensible to
also consider the standard deviation of {\tt vbroad} as measured
quantity and to check if its variance can be predicted by any of the
covariates. Indeed, aside from being caused by noise, it may partially
represent the time-dependence of the line broadening.
From our regression
analyses (presented in Appendix\,B)
we conclude that the noise contribution to the
standard deviation of {\tt vbroad} is dominant {  over intrinsic
  line-profile variability for the $\gamma\,$Dor pulsators.
For the SPB stars, the standard deviation of {\tt vbroad}
does relate to their surface rotation, 
effective temperature, and g-mode
frequency at the level of 20\% variance
reduction for the regression model based on these three covariates.}

Finally, we conclude that our analyses of $\sim\!15\,000$ new Gaia DR3 g-mode
pulsators bring the qualitative results on {\tt vbroad} by
\citet{Fremat2022} in full agreement with our quantitative assessments
on macroturbulence in g-mode pulsators, as already suggested 
by the simulation study in \citet{Aerts2009}.

\begin{acknowledgements}
  The research leading to these results has received funding from the
  KU\,Leuven Research Council (grant C16/18/005: PARADISE). CA and JDR
  acknowledge support from the BELgian federal Science Policy Office
  (BELSPO) through a PRODEX grant for the ESA space mission Gaia.  CA
  acknowleges financial support from the Research Foundation Flanders
  under grant K802922N (Sabbatical leave).  CA and GM are grateful for
  the kind hospitality offered by the staff of the Center for
  Computational Astrophysics at the Flatiron Institute of the Simons
  Foundation in New York City during their work visit in the fall of
  2022. The authors thank
{  the referee for the suggestion to investigate the sensitivity of
  the results to the temperature treshold used to reclassify the g-mode
  pulsators.  They also acknowledge}
  Dominic Bowman and Andrew Tkachenko for
  valuable comments which helped to improve the manuscript.
\end{acknowledgements}

\bibliographystyle{aa} 
\bibliography{Aerts-Gaia-gmode-pulsators-arxiv.bib}

\appendix

\section{Plots of the predictors for   {\tt vsini$_{-}$esphs}}
\begin{figure*}
\begin{center} 
\rotatebox{270}{\resizebox{6.1cm}{!}{\includegraphics{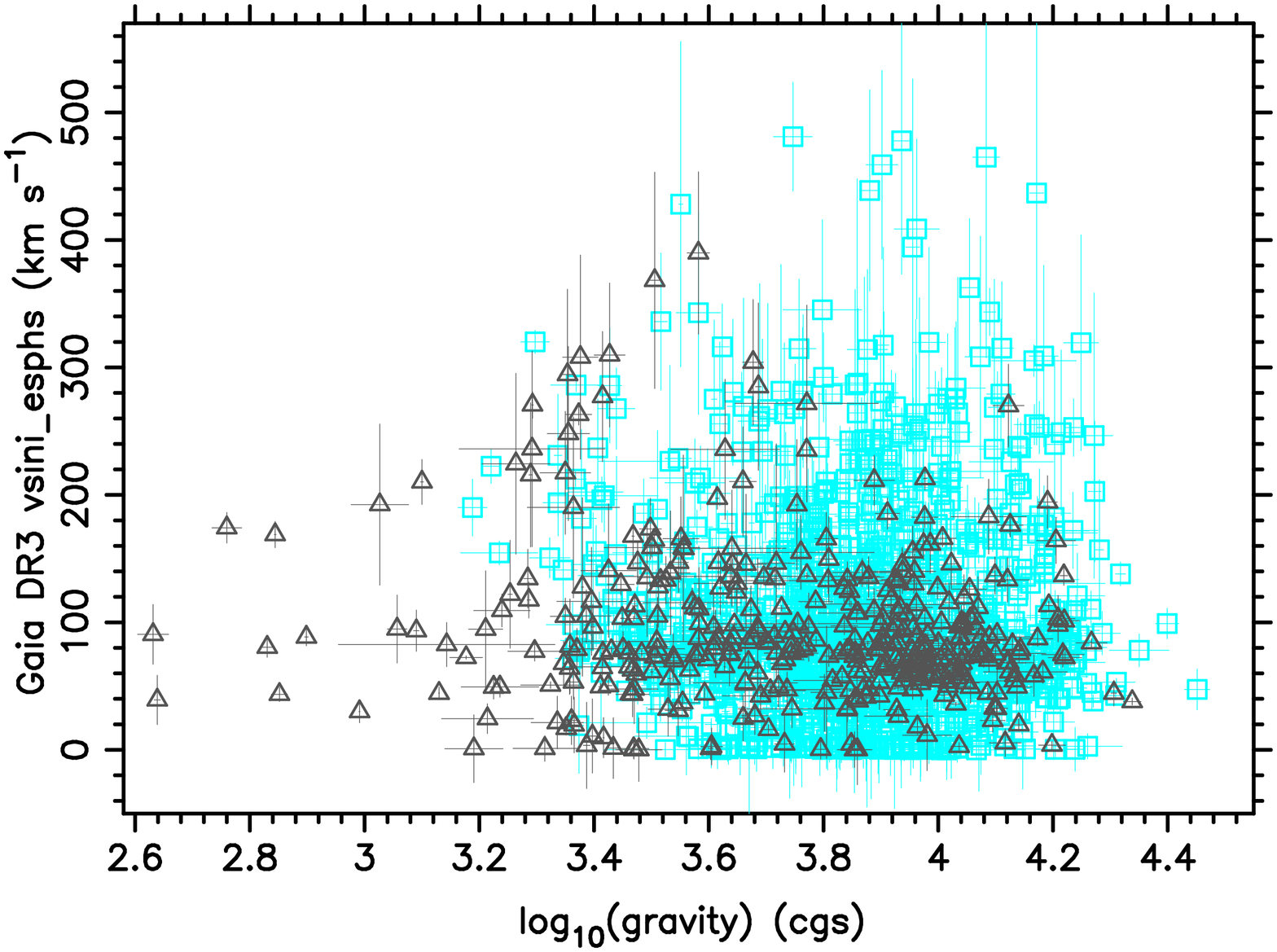}}}\hspace{0.5cm}
\rotatebox{270}{\resizebox{6.1cm}{!}{\includegraphics{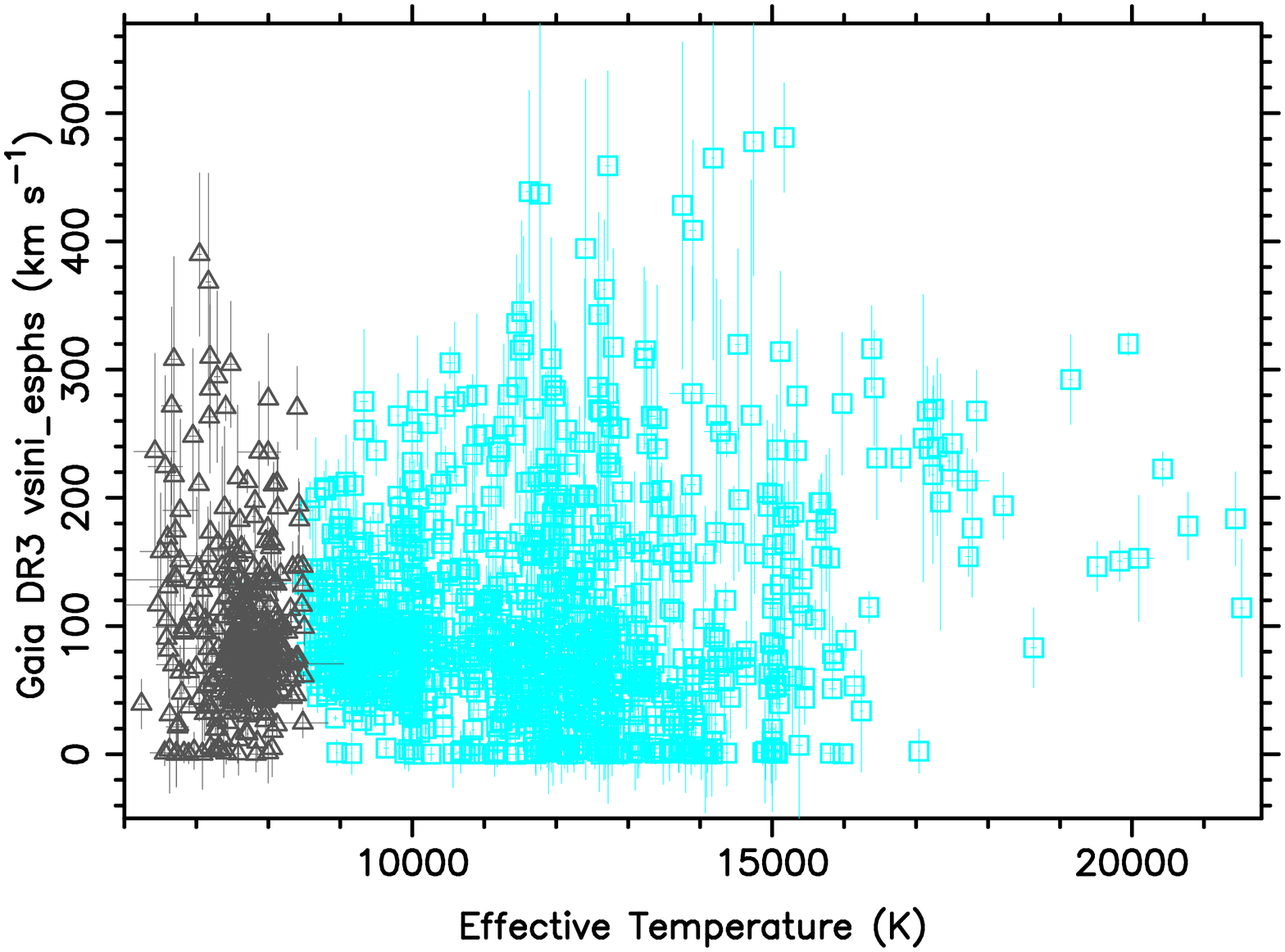}}}\vspace{0.5cm}
\rotatebox{270}{\resizebox{6.1cm}{!}{\includegraphics{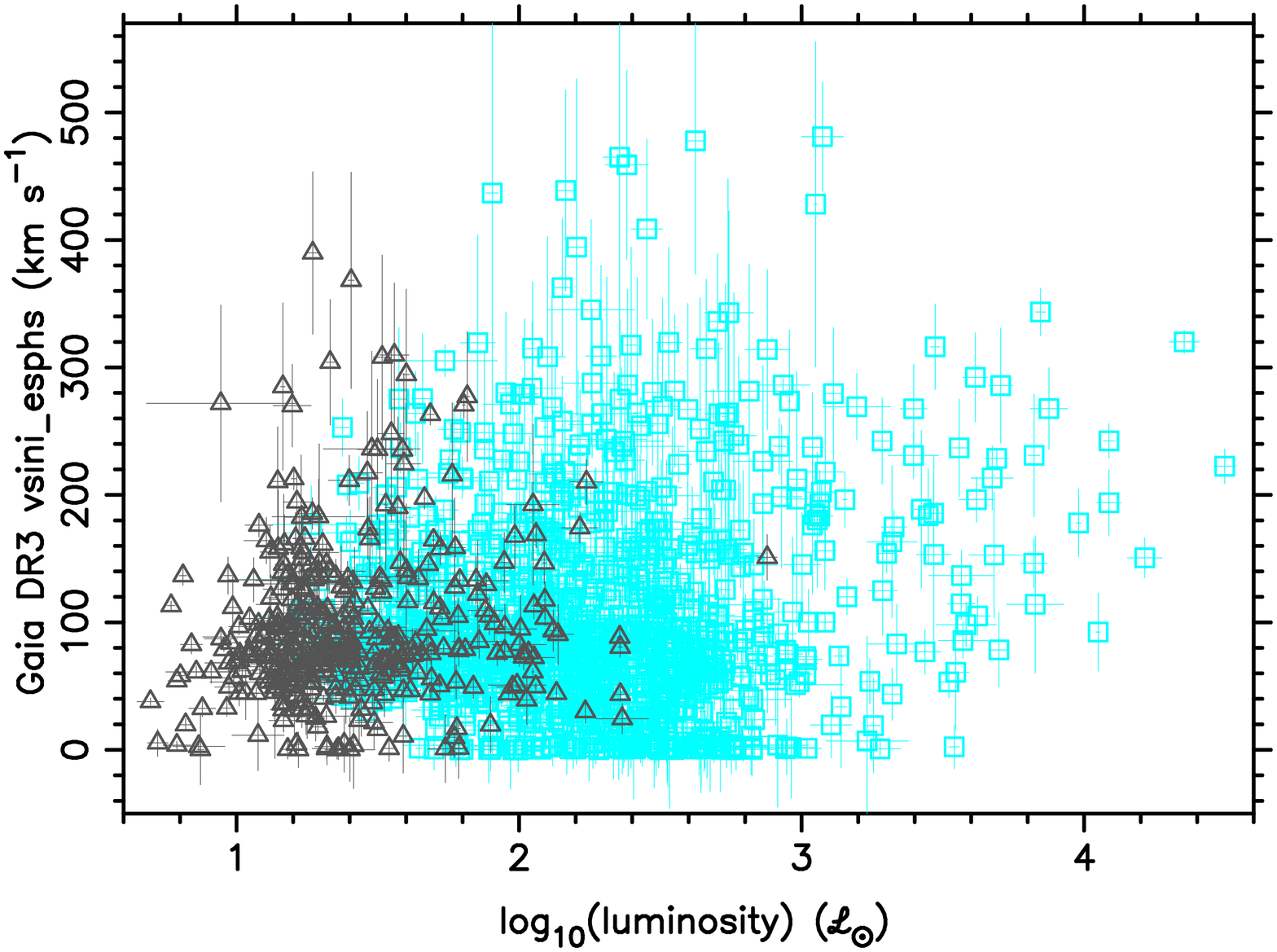}}}\hspace{0.5cm}
\rotatebox{270}{\resizebox{6.1cm}{!}{\includegraphics{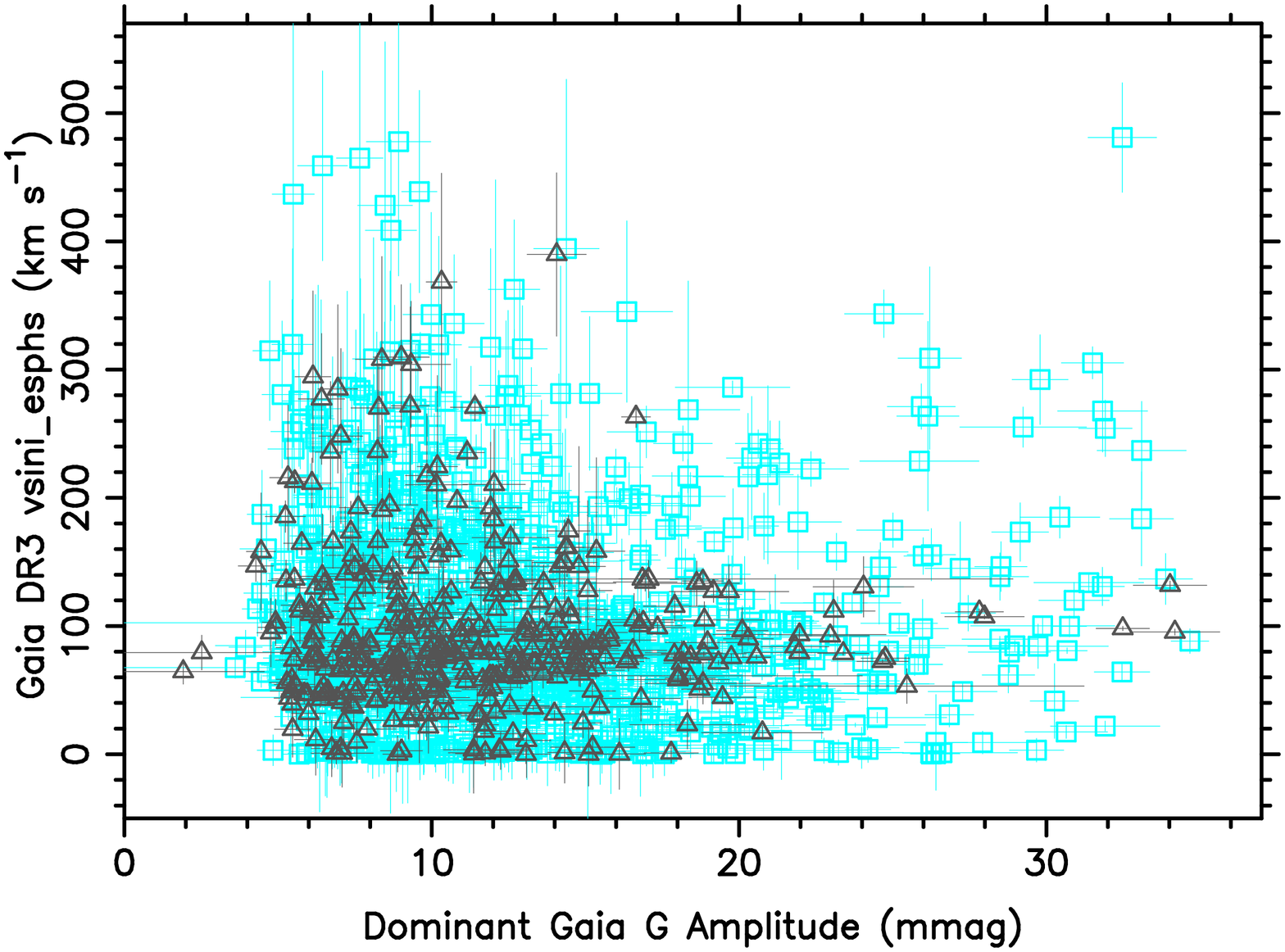}}}\vspace{0.5cm}
\rotatebox{270}{\resizebox{6.1cm}{!}{\includegraphics{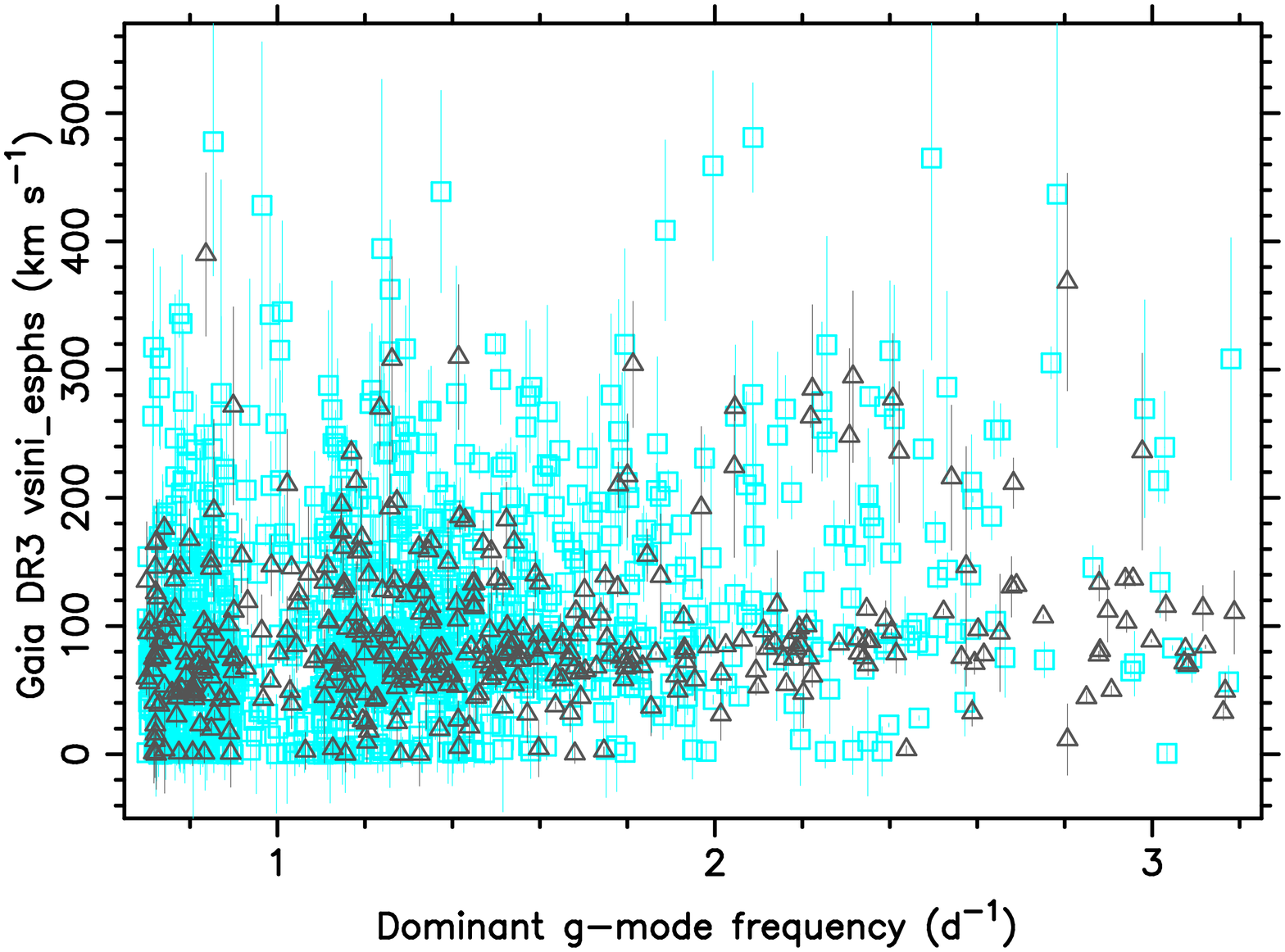}}}\hspace{0.5cm}
\rotatebox{270}{\resizebox{6.1cm}{!}{\includegraphics{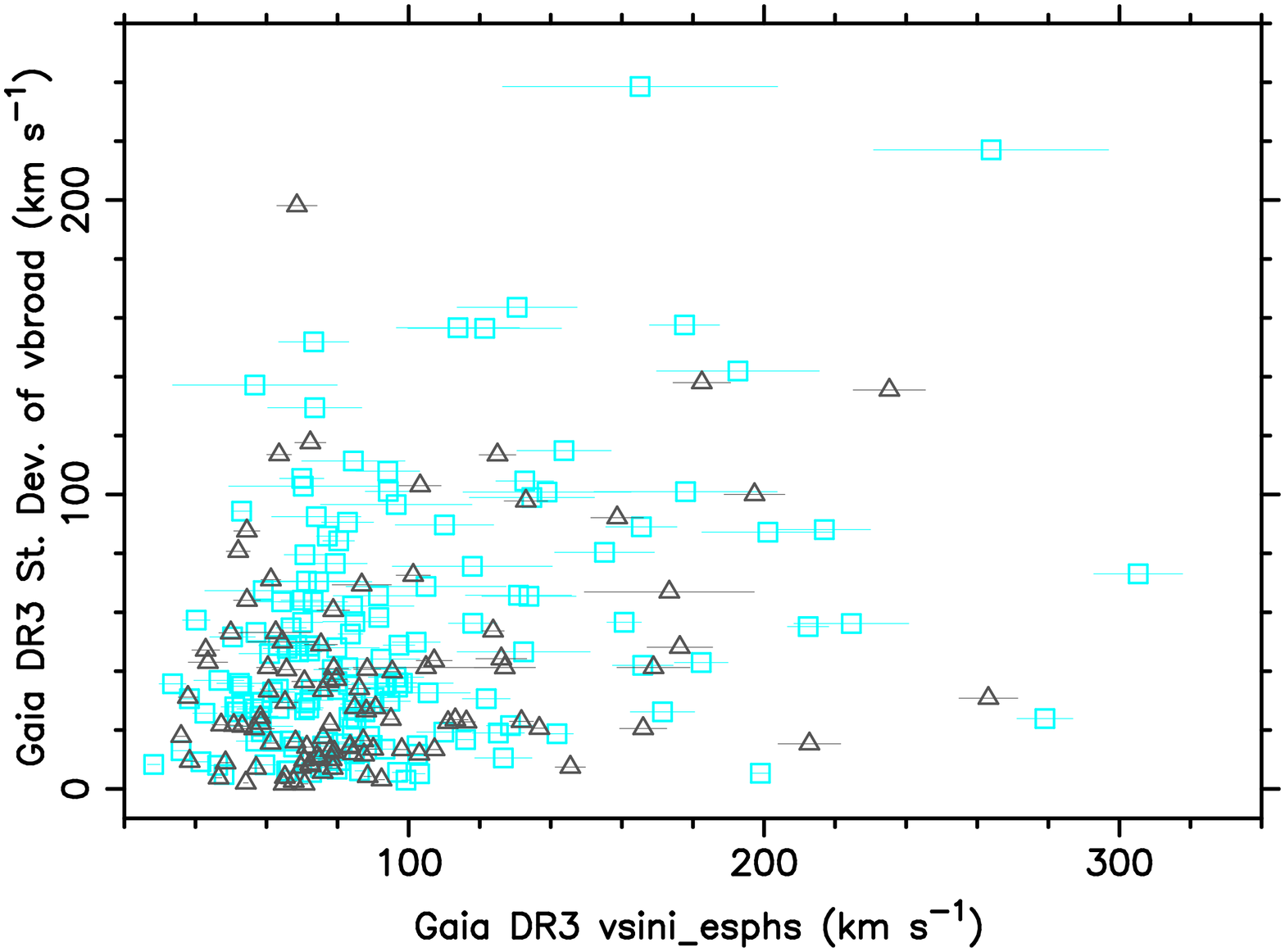}}}
\end{center}
\caption{\label{covariatesvsini} Gaia DR3 measurements of {\tt
    vsini$_{-}$esphs}
    versus each of the five covariates as indicated for the {  384}
  $\gamma\,$Dor (grey triangles) and {  1104} SPB (cyan squares) stars
  having these quantities available. The lower right panel shows the
  standard deviation of {\tt vbroad} as a function of
{\tt vsini$_{-}$esphs} for the {  100} $\gamma\,$Dor and {  190} SPB stars
having these quantities available. When invisible, the errors are smaller than the
  symbol sizes.}
\end{figure*}

\section{Regression models for the standard deviation of {\tt vbroad}}

The quantity {\tt vbroad} can be considered a simplified estimate of
the time-dependent second moment of a line profile
\citep{Aerts1992,DePauw1993,Aerts1994,BriquetAerts2003}.  It is
therefore meaningful to investigate if the standard deviation of {\tt
  vbroad} contains information on the time-dependent line-profile
variability of the Gaia DR3 g-mode pulsators.

In order to compute backward selection regression models for this
quantity, we now add {\tt vsini$_{-}$esphs} as sixth covariate (with
$\beta_6$ as notation for its regression coefficient). In line with
\citet{SSD2014} and 
\citet{Aerts2014b}, \citet{Nadya2022} have found from simulations that
the projected rotational velocity should first be derived from
high-resolution spectra before any meaningful derivation of left-over
additional time-dependent spectral line broadening caused by
tangential velocity fields can be done.  This is in agreement with the
methodology used by \citet{Aerts2014b} to assess the quality of
estimates for time-dependent macroturbulent broadening from the line's
moment variations in well-known B-type pulsators.

\begin{table}[h!]
  \tabcolsep=3.5pt
\begin{center}
  \caption{Estimates (standard errors) for the parameters of the
    errors-in-variables model for the standard deviation of {\tt
      vbroad}, measured for the {  1775} Gaia DR3 $\gamma\,$Dor
    stars. The results are based on backward selection considering
    a set of 6 (top) or 5 (bottom) predictors.
  \label{sigvbroadG02}}
\begin{tabular}{lcrrr}
\hline
Effect&Par.&Estimate (s.e.)&$p$-value&95\% conf. int.\\
\hline
\multicolumn{5}{c}{With {\tt vsini$_{-}$esphs} --- $R^2=0.09$}\\
\hline
Intercept            &$\beta_0$&$-$459(146)&&[$-$750;$-$169]\\
$\log\,g$                 &$\beta_2$&93(29)&0.0019&[35;151]\\
$\log(L/{\cal L}_\odot)$   &$\beta_3$&85(27)&0.0019&[32;138]\\
{\tt   vsini$_{-}$esphs}  &$\beta_6$&0.26(0.08)&0.0018&[0.10;0.42]\\
Res.\ var.           &$\sigma^2$&1122(166)&$<$0.0001&[793;1452]\\
\hline
\multicolumn{5}{c}{Without {\tt vsini$_{-}$esphs} --- $R^2=0.06$}\\
\hline
Intercept            &$\beta_0$&$-$517(135)&&[$-$781;$-$253]\\
$\log\,T_{\rm eff}$     &$\beta_1$&136(35)&0.0001&[66;205]\\
$\log(L/{\cal L}_\odot)$   &$\beta_3$&11(3)&$<$0.0001&[6;16]\\
$\nu$                 &$\beta_4$&8(1)&$<$0.0001&[6;10]\\
$A_\nu$             &$\beta_5$&$-$200(88)&0.0234&[$-$374;$-$27]\\
Res.\ var.           &$\sigma^2$&497(17)&$<$0.0001&[464;529]\\
\hline
\end{tabular}
\end{center}
\end{table}

\begin{table}[h!]
  \tabcolsep=3.5pt
\begin{center}
  \caption{Estimates (standard errors) for the parameters of the
    errors-in-variables model for the standard deviation of {\tt
      vbroad}, measured for the {  190} Gaia DR3 SPB pulsators.
    The results are based on backward selection considering a set of 6
    (top) or 5 (bottom) predictors.
  \label{sigvbroadG03}}
\begin{tabular}{lcrrr}
\hline
Effect&Par.&Estimate (s.e.)&$p$-value&95\% conf. int.\\
\hline
\multicolumn{5}{c}{With {\tt   vsini$_{-}$esphs} --- $R^2=0.20$}\\
\hline
Intercept            &$\beta_0$&$-$515(246)&&[$-$1001;$-$30]\\
$\log\,T_{\rm eff}$     &$\beta_1$&138(61)&0.0256&[17;258]\\
$\nu$                 &$\beta_4$&$-$17(6)&0.0031&[$-$28;$-$6]\\
{\tt vsini$_{-}$esphs}  &$\beta_6$&0.39(0.06)&$<$0.0001&[0.26;0.52]\\
Res.\ var.           &$\sigma^2$&1324(156)&$<$0.0001&[1016;1631]\\
\hline
\multicolumn{5}{c}{Without  {\tt   vsini$_{-}$esphs} --- $R^2=0.01$}\\
\hline
Intercept            &$\beta_0$&$-$515(237)&&[$-$382;$-$48]\\
$\log\,T_{\rm eff}$     &$\beta_1$&140(59)&0.0185&[24;256]\\
Res.\ var.           &$\sigma^2$&1641(169)&$<$0.0001&[1308;1974]\\
\hline
\end{tabular}
\end{center}
\end{table}

Table~\ref{sigvbroadG02} displays the regression models for the
standard deviation of {\tt vbroad} for the sample of {  1775} Gaia DR3
$\gamma$\,Dor stars. We perform the
backward selection twice, once with the covariate {\tt
  vsini$_{-}$esphs} and once without it.  For the model including it
as a potential predictor, we remove the insignificant contributions of
the variables
{ 
$\log\,g$, $A_\nu$, and $\log(L/{\cal L}_\odot)$}
(in that order). Not suprisingly, {\tt vsini$_{-}$esphs} is a
significant predictor for the standard deviation of {\tt vbroad},
along with the
{  gravity and luminosity} of the star.  For the regression
model without {\tt vsini$_{-}$esphs},
{  
only $\log\,g$
is removed and the predictive power decreases from 9\%} to 6\%
of the original variance.

The combined results in Tables\,\ref{vbroadgvsinig}, \ref{vbroadGcov},
and \ref{sigvbroadG02} reveal that rotational broadening of
$\gamma\,$Dor stars dominates the measurement of {\tt vbroad},
while the evolutionary status (by means of the
gravity and/or effective temperature) lies at the basis of the
significant predictors of
{\tt vbroad}.
{  
  Regression models for the standard deviation
have low predictive power and  are thus
harder to interpret, in line with} 
the caveats due to the systematic uncertainties connected
with the treatments of microturbulent broadening, $\log\,g$, and
$T_{\rm eff}$ highlighted by \citet{Tkachenko2020}. 

As for the SPB pulsators, Table~\ref{sigvbroadG03} displays the
regression models for the standard deviation of {\tt vbroad} from
backward selection for the {  190} SPB stars with such measurements.  For
the model with the covariate {\tt vsini$_{-}$esphs} included as a
potential predictor, the order of variables to be removed was
{ 
$\log\,g$, $A_\nu$, and 
  $\log(L/{\cal L}_\odot)$.}
The model explains {  20\%} of the variance in the standard
deviation of {\tt vbroad} and has {\tt vsini$_{-}$esphs},
$\log\,T_{\rm eff}$, and $\nu$ as
significant predictors.  For the model without {\tt
  vsini$_{-}$esphs} as a potential predictor,
{  all covariates aside from $\log\,T_{\rm eff}$
  were
removed as insignificant and the
predictive power disappears.} 
Thus, we
find that the standard deviation of the {\tt vbroad} measurements of
SPB stars {  is affected by their surface rotation and is
  connected to their effective temperature and 
dominant g-mode frequency.}

The regression models for the standard deviation of {\tt vbroad} have
low {  (SPB stars) to almost no ($\gamma\,$Dor stars)}
predictive power. This is not surprising as this standard
deviation is, at best, just one single value capturing the complicated
overall time-dependent broadening determined from only a few
scans across Gaia's field of view.
Our findings suggest that the noise contribution to the
standard deviation of {\tt vbroad} dominates over the one due to the
{  astrophysical parameters for the $\gamma\,$Dor stars.}

\end{document}